\documentclass[twocolumn]{aastex63}

\usepackage{CJKutf8}
    
\usepackage{graphics}                                           
\usepackage{color}     
\usepackage{amsfonts}
\usepackage{lipsum}
\usepackage{mathrsfs}
\usepackage{float}
\usepackage{tabularx}
\usepackage{mathtools}
\usepackage{cellspace,tabularx}
\usepackage{array}
\usepackage{multirow}
\usepackage{tcolorbox}

\usepackage{pifont}

\makeatletter
\newcommand{\skipitems}[1]{%
  \addtocounter{\@enumctr}{#1}%
}
\makeatother

\shorttitle{Cosmological Parameters from Momentum PS}
\shortauthors{Appleby et al.}

\begin{document}
\title{Cosmological Parameter Constraints from the SDSS Density and Momentum Power Spectra}

\author[0000-0001-8227-9516]{Stephen Appleby} 
\affiliation{Asia Pacific Center for Theoretical Physics, Pohang, 37673, Korea}
\affiliation{Department of Physics, POSTECH, Pohang, 37673, Korea}
\author[0000-0001-6400-1692]{Motonari Tonegawa} 
\affiliation{Asia Pacific Center for Theoretical Physics, Pohang, 37673, Korea}
\email{motonari.tonegawa@apctp.org}
\author[0000-0001-9521-6397]{Changbom Park}
\affiliation{School of Physics, Korea Institute for Advanced Study, 85
Hoegiro, Dongdaemun-gu, Seoul, 02455, Korea}
\author[0000-0003-4923-8485]{Sungwook E. Hong} 
\affiliation{Korea Astronomy and Space Science Institute,
776 Daedeok-daero, Yuseong-gu, Daejeon, 34055, Republic of Korea}
\affiliation{Astronomy Campus, University of Science and Technology,
776 Daedeok-daero, Yuseong-gu, Daejeon, 34055, Republic of Korea}
\author[0000-0002-4391-2275]{Juhan Kim} 
\affiliation{Center for Advanced Computation, Korea Institute for Advanced Study, 85 Hoegiro, Dongdaemun-gu, Seoul, 02455, Korea}
\author[0000-0003-0134-8968]{Yongmin Yoon} 
\affiliation{Korea Astronomy and Space Science Institute,
776 Daedeok-daero, Yuseong-gu, Daejeon, 34055, Republic of Korea}

%%%%%%%%%%%%%%%%%%%%%%%%%%%%%%%%%%%%%%%%%%%%%%%%%%%%%%%%%%%%%%%%%%%             
\begin{abstract}
We extract the galaxy density and momentum power spectra from a subset of early-type galaxies
in the SDSS DR7 main galaxy catalog. Using galaxy distance information inferred from the improved fundamental plane described in \citet{Yoon_2020}, we reconstruct the peculiar velocities of the galaxies and generate number density and density-weighted velocity fields, from which we extract the galaxy density and momentum power spectra. We compare the measured values to the theoretical expectation of the same statistics, assuming an input $\Lambda$CDM model and using a third-order perturbative expansion. After validating our analysis pipeline with a series of mock data sets, we apply our methodology to the SDSS data and arrive at constraints $f\sigma_{8} = 0.471_{-0.080}^{+0.077}$ and $b_{1}\sigma_{8} = 0.920_{-0.070}^{+0.070}$ at a mean redshift $\bar{z} = 0.04$. Our result is consistent with the Planck cosmological best fit parameters for the $\Lambda$CDM model. The momentum power spectrum is found to be strongly contaminated by small scale velocity dispersion, which suppresses power by $\sim {\cal O}(30\%)$ on intermediate scales $k \sim 0.05 \, h \, {\rm Mpc}^{-1}$. 
\end{abstract}


%\pacs{98.80.-k,98.80.Bp}

%\maketitle
%%%%%%%%%%%%%%%%%%%%%%%%%%%%%%%%%%%%%%%%%%%%%%%%%%%%%%%%%%%%%%%%%%%%%%%%%%%

\section{Introduction}

Extracting information from the spatial distribution of galaxies is a perpetual occupation within the cosmological community. From angular positions and redshifts, one can reconstruct the galaxy number density field and hence measure the galaxy $N$-point statistics in position and Fourier space. Typically the two-point correlation function and power spectrum are utilised, owing to their relative simplicity and considerable constraining power \citep{Blake:2011,Anderson:2014,Oka:2014,Okumura:2016,delaTorre:2017}. Previously intractable higher point statistics are increasingly studied due to rapid advances in numerical statistical modelling and larger data volumes \citep{Gil-Martin:2017,Yankelevich:2019,Philcox:2022}. Alternative approaches to the standard $N$-point functions are also progressively being explored \citep{Pisani:2015,Pan:2019vky,Uhlemann:2020,Navarro:2021,Appleby:2021xoz,Qin:2023dew}.

Galaxy properties are measured in redshift space, in which cosmological redshifts are contaminated by local peculiar velocities parallel to the line of sight. By modelling this redshift-distortion effect we can constrain not only the parameters governing the shape and amplitude of the power spectrum, but also the rate at which structures are forming. The reason is that peculiar velocities trace in-fall into gravitational potentials, a phenomenon that occurs on all scales \citep{Kaiser:1987}. Unfortunately, the effect of redshift space distortion is strongly degenerate with the overall amplitude of the power spectrum -- at the level of linearized perturbations the two are exactly degenerate. However, if we complement the galaxy positional information by also measuring their velocities, then we can additionally extract the velocity power spectrum and break this degeneracy, placing simultaneous constraints on the amplitude of the galaxy power spectrum and the growth rate of structure. By modelling the quasi-linear scales using higher order perturbation theory, one can obtain additional information on galaxy bias and improved constraints on cosmological parameters \citep{dAmico:2020}. On small scales, stochastic velocities arising from bound structures dominate the cosmological signal, a phenomenon known as the Finger-of-God effect \citep{Jackson:1971sky,1994ApJ...431..569P,1995ApJ...448..494F,Juszkiewicz:1998em,Hikage:2013yja,Tonegawa:2020wuh,PhysRevD.70.083007,10.1111/j.1365-2966.2010.17581.x,Jennings_2010,Okumura_2010,Kwan_2012,10.1093/mnras/stu2460}.

Distance measurements to galaxies are difficult to infer, as they require prerequisites for a physical scaling relation that can be used to convert dimensionless quantities such as redshift into distance. The complicated dynamics and evolution of galaxies means that there is no simple universal scale associated with their morphology, although certain subsets are known to be reasonably well approximated as virialised systems. For such subsets, we are able to directly measure the distance to recover the velocity field of the matter distribution.

The velocity field is problematic to reconstruct since it is expected to be non-zero even in empty spaces such as voids\footnote{In spite of this, the velocity field has been extensively studied in the literature, see for example \citep{Davis:1982gc,McDonald:2008sh,Kim:2019kls,Howlett:2016urc,Adams:2020dzw,Johnson:2014kaa,Howlett:2017asq,Adams:2017val,Koda:2013eya}.}. To evade this problem, the density weighted velocity field -- or momentum field -- was first proposed as a more amenable statistic in \citet{Park:2000rc}, since it naturally approaches zero in regions where the density is low. This statistic was used in \citet{Park:2005bu} to place cosmological parameter constraints with some nascent large scale structure catalogs, but then the idea lay dormant for a time, before being picked up in a series of recent works \citep{Howlett:2019bky,Qin:2019axr}. Significant improvements towards understanding the quasi-linear ensemble average of the momentum field has been made in the intervening period \citep{Vlah:2012ni,Saito:2014qha}.

In this work we take the redshift and velocity information of a subset of SDSS galaxies measured in \citet{Yoon_2020}, extract the galaxy density and momentum power spectra and fit the standard cosmological model to the statistics, inferring a constraint on the galaxy power spectrum amplitude $b_{1}\sigma_{8}$ and growth rate $f\sigma_{8}$. While working towards this goal, we encounter a series of difficulties that we document in the following sections, pertaining to the convolution of the mask, residual parameter degeneracies and the effect of non-linear velocity dispersion on the shape of the momentum power spectrum. We highlight the assumptions and approximations required to arrive at our parameter constraints throughout.

In Section \ref{sec:II_data} we review the data used in our analysis, and the method that we use to extract the power spectra from the point distribution. We present the ensemble expectation values of the statistics, to which we compare our measurements in Section \ref{sec:theory} including the effect of the mask. Some preliminary measurements from mock galaxy catalogs are presented in Section \ref{sec:mocks}. Our main results can be found in Section \ref{sec:results}, where we provide measurements of the power spectra from the data, and the resulting cosmological parameter constraints are in Section \ref{sec:parm_est}. We discuss our findings in Section \ref{sec:disc} and review the assumptions made in arriving at our  results.

\section{Data} 
\label{sec:II_data}
\begin{figure*}
 \begin{center}
  \includegraphics[width=0.9\textwidth]{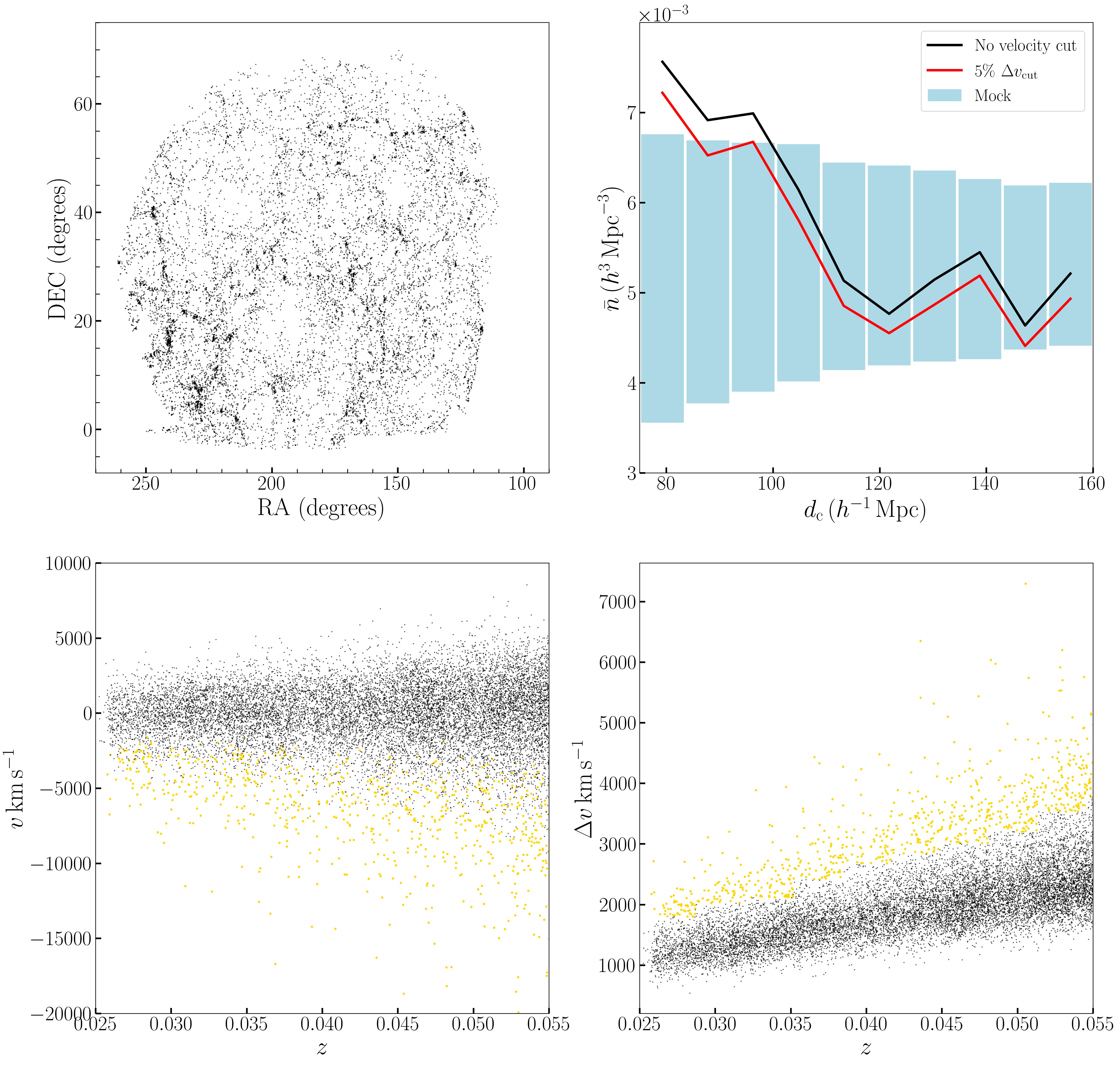}
 \end{center}
 \caption{\label{fig:mom5} From the SDSS FP catalog, the angular distribution of ETGs used in this work (top left), the number density as a function of comoving distance with (red) and without (black) a velocity uncertainty cut (top right). The blue points/errorbars in the top right panel are the mean and standard deviation of the mock catalogs, discussed further in Section \ref{sec:mocks}. In the lower panels we present measured galaxy velocity $v$ against redshift (lower left panel) and velocity measurement uncertainty $\Delta v$ against redshift (lower right panel). The black points in the lower panels represent the galaxies used in our analysis, and the yellow points those that are cut by our choice of velocity uncertainty limit. The uncertainty in the velocity measurement $\Delta v$ is negatively correlated with the measured velocity $v$, hence the $\Delta v$ cut removes the large negative velocity galaxies (cf. lower left panel).
}
\end{figure*}

A subset of the SDSS Data Release 7 \citep[DR7;][]{Abazajian:2009} classified as early-type galaxies (ETGs) in the KIAS value-added galaxy catalog (VAGC; \citet{Choi:2007}) is used to measure the galaxy density and momentum power spectra. It is based on the New York University Value-Added Galaxy Catalog \citep{Blanton:2005}, where missing redshifts are supplemented from various other redshift catalogs to improve the completeness at higher redshift. The classification between early and late type is based on $u-r$ colour, $g-i$ colour gradient and inverse concentration index in the $i$-band \citep{Park:2005zb,Choi:2006qg}. The result of the automated classification is corrected by visual inspection; we direct the reader to \citet{Park:2005zb} for details. We use a sub-sample obtained by applying redshift and magnitude cuts. ETGs are selected in the redshift range $0.025 \leq z_{\rm spec} < 0.055$, with the lower limit corresponding to a distance $\sim 75 \, h^{-1} \, {\rm Mpc}$ to mitigate large peculiar velocity effects in the local Universe. A de Vaucouleur absolute magnitude cut of $M^{\rm dV}_{\rm r} \leq -19.5$ is also applied. There are a total of $16,283$ galaxies in the sample, although we make further cuts below. In the top left panel of Figure %\ref{fig:1}
\ref{fig:mom5} we present the angular distribution of the ETGs on the sky. We have removed the stripes from the data, focusing only on the largest contiguous SDSS footprint region.

ETGs are selected since they are bulge-dominated systems with velocity dispersions well described using the virial theorem. As a result, ETGs lie on the fundamental plane (FP) in the space of three-dimensional variables

\begin{equation} \log_{10}R_{e} = a \log_{10}\sigma_{0} + b \mu_{e} + c \end{equation} 

\noindent where $\sigma_{0}$ is the central velocity dispersion, and $\mu_{e}$ is the mean surface brightness within half-light radius $R_{e}$. In principle, $a$ and $b$ can be deduced from the virial theorem, but practically $a$, $b$, $c$ are empirically determined parameters. In \citet{Yoon_2020} it was noted that a subsample of ETG galaxies exhibit a significantly smaller scatter on the FP plane and were selected to generate a catalog with smaller intrinsic scatter in the velocity reconstruction.  Specifically, the FP of old ETGs with age$\gtrsim9$ Gyr has a smaller scatter of $\sim0.06$ dex ($\sim14\%$ in the linear scale) than that of relatively young ETGs with age $\lesssim6$ Gyr that exhibits a larger scatter of $\sim0.075$ dex ($\sim17\%)$. For the subsample of young ETGs, less compact ETGs have a smaller scatter on the FP ($\sim0.065$ dex; $\sim15\%$) than more compact ones ($\sim0.10$ dex; $\sim23\%$). By contrast, the scatter on the FP of old ETGs does not depend on the compactness of galaxy structure. The use of the FPs with smaller scatters allows for more precise distance measurements,\footnote{The distance of a galaxy is determined by comparing its angular size with the expected physical size derived from the FP.} which in turn facilitates a more accurate determination of the peculiar velocities of ETGs.

From the sample, we exclude galaxies with a large velocity uncertainty $\Delta v$. In the lower right panel of Figure \ref{fig:1} we present the velocity uncertainty of each galaxy in the SDSS FP catalog as a function of redshift. We observe a systematic increase in velocity uncertainty as a function of $z$, which is typical in velocity measurements. The galaxies with extreme velocity uncertainties will generate a large, non-Gaussian noise  contribution to the momentum power spectrum that is undesirable for our purposes. For this reason, we generate a Constrained B-spline fit in the $\Delta v$ - $z$ plane \citep{2015ascl.soft05010N, 2017ApJ...847..134K, 2018AJ....155..137B, 2019ApJ...872...63L}, and then cut all galaxies in the sample above the $95\%$ percentile of the velocity uncertainty. The yellow points are those that are cut using this procedure, which constitute a total of $734$ objects. This is approximately $5\%$ of the $14,180$ FP galaxies that lie in the range $0.025 \leq z \leq 0.055$ and within the main SDSS footprint. In the bottom left panel of Figure \ref{fig:1} we present the velocities of each galaxy as a function of redshift. The gold points are those cut by our procedure; we observe that those objects with a large negative velocity are removed. The velocity distribution is skewed towards negative values because galaxies are more likely to be scattered into the redshift range that out. Our choice largely removes this systematic artifact. Note that one could instead choose to weight each galaxy with some function their velocity uncertainty when constructing the momentum field \citep{Qin:2019axr}. This achieves practically the same goal.

%From the sample, we exclude all galaxies with an observed peculiar velocity larger than $|v_{\rm pec}| = 5000 \, {\rm km \, s^{-1}}$. In Figure \ref{fig:mom5} we present the velocity of each galaxy as a function of redshift (bottom left panel) and the measurement error $\Delta v$ against velocity $v$ (bottom right panel). In both panels we observe a skewed distribution of velocities, with a preference for large, negative values. The skewness arises because galaxies that are scattered with large, positive velocity will be pushed outside the redshift range of the galaxy catalog, whereas those with negative velocities will enter the catalog from above the redshift limit. This effect is asymmetric since we measure galaxies in a cone of increasing volume with redshift -- there are more galaxies at high redshift that can be scattered towards the observer. We apply the quoted velocity cut -- presented as %black red dashes in the Figure. This choice also removes the majority of galaxies with large velocity uncertainties. After applying the cut, the number of galaxies in the catalog is $N_{\rm gal} = 15,442$ and the mean redshift is $\bar{z} = 0.043$. 

Making a velocity selection may introduce an artificial gradient in the number density, and this must be checked. In Figure \ref{fig:mom5} we present the number density of the SDSS FP catalog used in this work, as a function of comoving distance (top right panel). The black line is the number density without a cut, and the red line is the sample with the velocity uncertainty cut applied. The blue points/error bars are the sample mean and standard deviation of the number density of a set of mock samples, which are described in Section \ref{sec:mocks}. Our selection does not introduce any significant redshift gradient, only acting to reduce the overall number density by $\sim 5\%$. However, the effect is considerably less than the $1\sigma$ density variation in the mocks that we use in our analysis (cf. blue points/error bars). 

We also observe a large fluctuation in the number density of the full sample, with a higher number density of ETGs in the first three bins (cf. black line). We expect that this is a statistical fluctuation arising from the modest volume of the sample, because the number density of the full sample does not show a systematic gradient with distance. Rather, the first three bins show a $\sim 1\sigma$ high number density. We do not sub-sample the galaxies to remove this overdensity, on the grounds that statistical fluctuations should not be removed by pre-processing the data. This local overdensity is very likely due to the presence of the CfA great wall, spanning the distance of approximately $60$ to $110 \, h^{-1} \, {\rm Mpc}$ and consisting of rich galaxy clusters such as Coma, Leo, and many in the Hercules Superclusters. Regardless, the variation of the density is a $\sim 1\sigma$ effect, and is not unusual compared to individual mock samples. In Appendix C we present some $\bar{n}$-$z$ curves from a random sample of mocks, finding similar variation in number density with redshift.  The mean of the mock data presents no systematic evolution in number density with redshift, as expected.

\subsection{Power Spectrum Estimator} 

To estimate the power spectrum, we first bin the galaxy catalog into a regular lattice that we then Fourier transform. With this aim in mind, we generate a box of size $L = 512 \, h^{-1} \, {\rm Mpc}$ to enclose the data, and create a regular lattice of grid points with resolution $\Delta = 512/256 = 2 \, h^{-1} \, {\rm Mpc}$ per side. The observer is placed at the center of the box. All galaxies are assigned to the pixels according to the cloud-in-cell (CIC) algorithm, and for both the galaxy density and momentum fields each galaxy is inversely weighted by its KIAS %vagc
VAGC angular selection function weight\footnote{We do not apply any other weighting scheme to the data, such as the FKP weights \citep{Feldman:1994} since this quantity is optimized for a Gaussian random field. Although this might be a non-issue, we wish to minimize any assumptions on the data.}. We use the redshift of each galaxy corrected to the CMB frame, and assume a flat $\Lambda$CDM cosmology with parameters $\Omega_{m} = 0.3$,  $w_{\rm de} = -1$ to infer the comoving distance. Once all galaxies have been assigned to the grid, we apply the SDSS angular selection function as a binary HEALPix\footnote{http://healpix.sourceforge.net} mask of resolution $N_{\rm side} = 512$ \citep{Gorski:2004by}, zeroing any pixel if its angular weight is less than $w < 0.9$. We also apply a radial cut, and zero all pixels that lie $r \leq 80 h^{-1} \, {\rm Mpc}$ and $r \geq 160 h^{-1} \, {\rm Mpc}$ relative to the observer. The $\ell=0$ mode of the number density, or momentum, power spectrum is then given by \citep{Yamamoto:2006},

\begin{widetext}
\begin{equation}\label{eq:mom1}  |F_{\ell}(k)|^{2}  = {2 \ell +1 \over V} \int {d\Omega_{k} \over 4\pi} \int d^{3}r \int d^{3}r' F({\bf r}) F({\bf r'}) {\cal L}_{\ell}({\bf \hat{k}}\cdot{\bf \hat{r}'}) e^{i {\bf k} \cdot ({\bf r - r'})} ,
\end{equation}
\end{widetext}

\noindent where $F({\bf r})$ is constructed from the galaxy density $n({\bf r})$, its mean $\bar{n}$, and the line-of-sight velocity $u({\bf r})$ \citep{Howlett:2019bky},
%defined as 

\begin{eqnarray}
F^{\delta}({\bf r}) &=&  (n({\bf r}) - \bar{n})/\bar{n} , \\ 
\label{eq:fp} F^{p}({\bf r}) &=& n({\bf r}) u({\bf r}) ,
\end{eqnarray}

\noindent and ${\cal L}_{\ell}(x)$ are the Legendre polynomials.  We perform the integrals in Equation (\ref{eq:mom1}) as a double sum over unmasked pixels, which is a tractable procedure due to the limited volume of the data. Here, ${\bf r}$ and ${\bf r'}$ are vectors pointing to the pixels in the double sum. Using the Rayleigh expansion of a plane wave -- 

\begin{equation}  e^{i {\bf k} \cdot {\bf r}} = \sum_{\ell} i^{\ell} (2\ell + 1) j_{\ell}(kr) {\cal L}_{\ell}({\bf \hat{k}} \cdot {\bf \hat{r}}) ,
\end{equation}
where $j_\ell$ is the spherical Bessel function of $\ell$-th order,
%\noindent 
the integral over $\Omega_{k}$ reduces to 

\begin{equation} \int {d\Omega_{k} \over 4\pi} e^{i {\bf k} \cdot {\bf y}} {\cal L}_{L}({\bf \hat{k}} \cdot {\bf \hat{r}'}) = i^{L} j_{L}(ky) {\cal L}_{L}({\bf \hat{y}} \cdot {\bf \hat{r}'}) ,\end{equation} 

\noindent where ${\bf y} = {\bf r} - {\bf r'}$. We are only extracting the $L=0$ mode from the data, so take ${\cal L}_{L=0}({\bf \hat{y}} \cdot {\bf \hat{r}'}) = 1$. The contributions to the $\Omega_{k}$ integral from higher  $\ell$-modes in the Rayleigh expansion are negligible, consistent with noise. 

The FP peculiar velocity uncertainty of each galaxy is the dominant contaminant in the reconstruction of the momentum field, and in what follows we will treat the velocity uncertainty as a Gaussian white noise contribution to the momentum power spectrum. This is discussed further in Section \ref{sec:results}.

\section{Theoretical Density/Momentum Power Spectra}
\label{sec:theory}

We compare the measured power spectra to their ensemble averages, accounting for perturbative non-Gaussianity due to gravitational evolution. We now describe the theoretical model used in this work, which is derived elsewhere \citep{Vlah:2012ni,Saito:2014qha}. The theory and numerical implementation of the so-called distribution function approach can be found in \citet{Seljak:2011tx,McDonald:2009hs,Okumura:2011pb,Okumura:2012xh,Vlah:2012ni}, and other perturbation theory modelling treatises in \citet{Matsubara:2007wj, Bernardeau:2001qr, Pueblas:2008uv, Takahashi:2012em, Kwan:2011hr, Carlson:2012bu,Zheng:2016zxc}. 

We write the density of the matter distribution as $\rho({\bf x})$ and the corresponding cosmological velocity vector field as ${\bf v}({\bf x})$. The density can be decomposed into a time dependent average $\bar{\rho}(t)$ and fluctuations $\delta$ according to $\rho/\bar{\rho} = 1 + \delta({\bf x})$. The momentum field is defined as the density-weighted velocity ${\bf p}({\bf x}) = [ 1 + \delta({\bf x}) ] {\bf v}({\bf x})$ \citep{Park:2000rc}. We can only measure the radial component of the galaxy velocities, $p_{\parallel}= (1 + \delta) v_{\parallel}$ where $v_{\parallel} = {\bf v} \cdot {\bf \hat{e}}_{\parallel}$, and ${\bf \hat{e}}_{\parallel}$ is the unit vector pointing along the line of sight. 

We measure galaxy positions in redshift space. If we denote real and redshift space comoving distances as ${\bf r}$ and ${\bf s}$ respectively, they are related according to the relation 

\begin{equation}\label{eq:rsp} {\bf s} = {\bf r} + {1 \over aH} {\bf \hat{e}}_{\parallel}  ({\bf v} \cdot {\bf \hat{e}}_{\parallel}) ,\end{equation} 

\noindent where $a$ and $H$ are the scale factor and Hubble parameter, respectively. The density field in real ($\delta$) and redshift ($\tilde{\delta}$) space are correspondingly related according to \citet{Kaiser:1987}

\begin{equation} \tilde{\delta}({\bf x}) = \left[ 1 + f \left( {\partial^{2} \over \partial r^{2}} + {2 \over r} {\partial \over \partial r} \right) \nabla^{-2} \right]  \delta ({\bf x}) , 
\end{equation} 

\noindent where $f = d\log D/d\log a$ is the linear growth rate, and $D$ is the growth factor.
The quantity in the square bracket on the right hand side is the radial redshift space distortion operator. This relation is valid to linear order in the perturbations $\delta$ and ${\bf v}$. 

The plane parallel approximation is often used in redshift space analysis. It is an approximation in which a constant, common line of sight vector is assigned to every galaxy. This neglects the radial nature of redshift space distortion, and is appropriate in the limit in which the galaxy sample is far from the observer and localised to a patch in the sky. With this approximation, the redshift space density field can be written in simplified form in Fourier space as 

\begin{equation} \label{eq:df}  \tilde{\delta}(\mu, {\bf k}) = (1 + f \mu^{2}) \delta({\bf k}) , \end{equation}
where $\mu = k_{\parallel}/k$, and $k_{\parallel}$ is the component of the Fourier modes aligned with the line of sight vector, which is constant in the plane parallel limit\footnote{The density field in equation (\ref{eq:df}) cannot be physically realised without some window function convolution, as the condition implied by the plane parallel assumption -- localisation to a patch on the sky -- is not consistent with periodicity.}. This relation is valid in the large scale limit where the linear perturbation theory can be applied. The power spectrum of $\tilde{\delta}$ is 

\begin{equation} \grave{P}^{\tilde{\delta}}(\mu, k) = (1 + f \mu^{2})^{2} \grave{P}^{\delta}(k)  .
\end{equation}

Throughout this work we use an accent $\grave{}$ to distinguish theoretical power spectra from the same quantities measured from galaxy distributions. Similarly, in the same limits (plane parallel, linear perturbations), the momentum and velocity fields are equivalent and the $k_{\parallel}$ component of the velocity power spectrum can be approximated as 

\begin{equation} \grave{P}^{\tilde{p}_{\parallel}}(\mu,k) \simeq \mu^{2} f^{2} (aH)^{2} {\grave{P}^{\delta}(k) \over k^{2}}. 
\end{equation}

\begin{figure*}
 \begin{center}
  \includegraphics[width=0.9\textwidth]{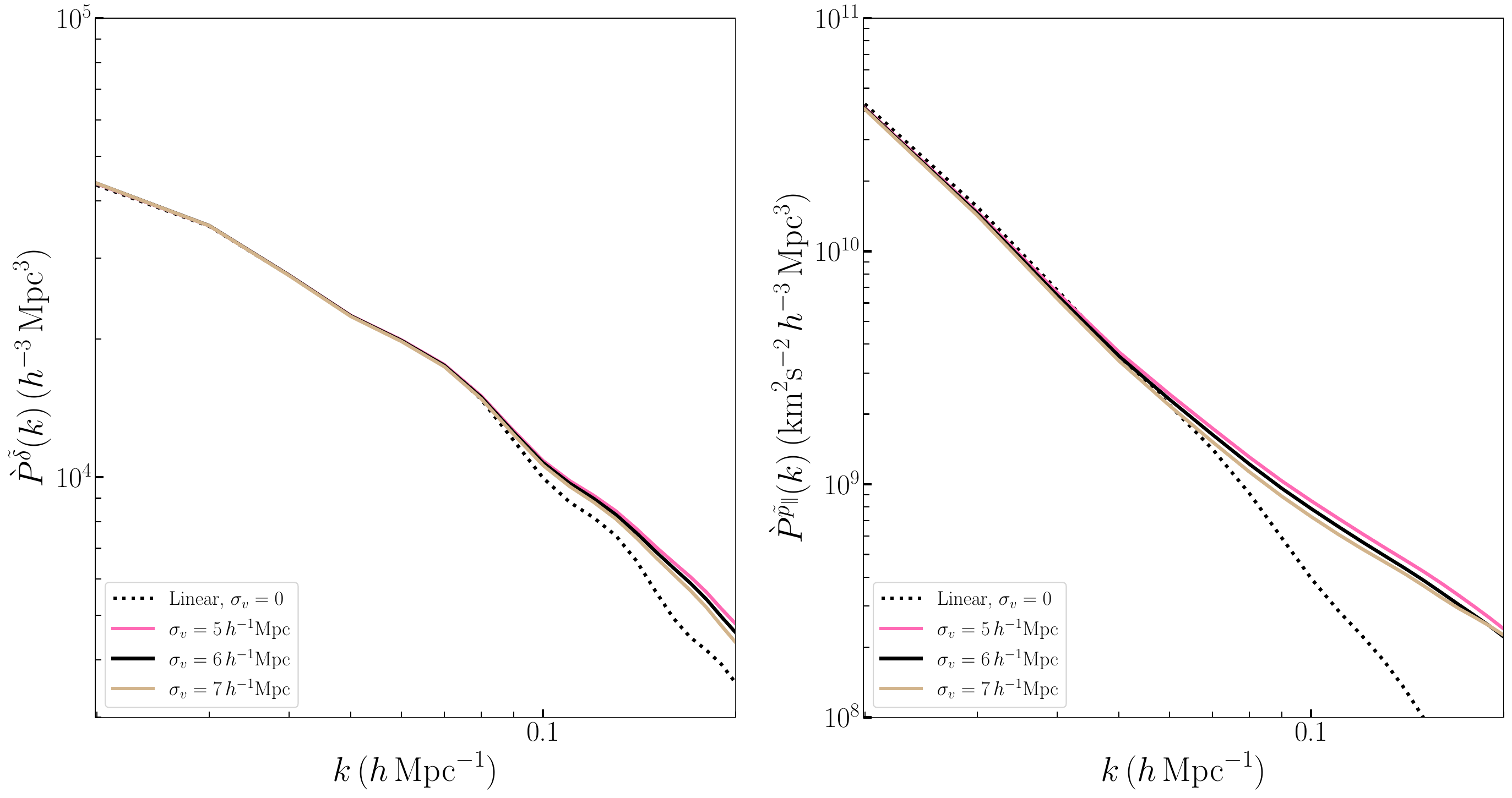}
 \end{center}
 \caption{\label{fig:theory} The non-linear theoretical power spectra $\grave{P}^{\tilde{\delta}}$ (left panel) and $\grave{P}^{\tilde{p}_{\parallel}}$ (right panel)
for different values $\sigma_{v} = (5,6,7) h^{-1} \, {\rm Mpc}$ (pink, black and tan lines). 
The black dashed lines are the corresponding linear power spectra, with $\sigma_{v} = 0$. 
}
\end{figure*}

Linear perturbation theory is not sufficient for the extraction of information on quasi-linear scales. Also, the relations above are valid for the dark matter field and the galaxy bias -- both linear and at higher order in the perturbations -- must be accounted for. In a series of works \citep{Vlah:2012ni,Saito:2014qha}, the galaxy density and momentum power spectra to third order in redshift space have been constructed. The form is neatly summarized in Appendix A of \citet{Howlett:2019bky}, and we direct the reader to that work and \citet{Vlah:2012ni} for the complete derivation. The power spectra can be written as 

\begin{widetext}
\begin{eqnarray}\label{eq:th1} & & \grave{P}^{\tilde{\delta}}(\mu, k) = P_{00} + \mu^{2} \left( 2P_{01} + P_{02} + P_{11} \right) + \mu^{4} \left( P_{03} + P_{04} + P_{12} + P_{13} + P_{22}/4 \right) , \\ 
\label{eq:th2} & & \grave{P}^{\tilde{p}_{\parallel}}(\mu,k) = (aH)^{2} k^{-2} \left[ P_{11} + \mu^{2} \left( 2P_{12} + 3P_{13} + P_{22} \right) \right] .
\end{eqnarray} 
\end{widetext}

\noindent We do not repeat the expressions for $P_{ij}$ in this work; they are combinations of convolutions of the linear matter power spectrum. Their exact functional form can be found in Appendix A of \citet{Howlett:2019bky} and Appendix D of \citet{Vlah:2012ni}.

In redshift space the power spectrum is no longer only a function of the magnitude of Fourier mode, due to the anisotropy generated along the line of sight\footnote{If we additionally drop the plane parallel approximation, then translational invariance is lost and the power spectrum becomes a function of the angle between line of sight and separation vector between tracers.}.  The standard approach is to decompose the power spectra into multipoles;

\begin{equation} \label{eq:psd} \grave{P}^{\tilde{\delta}, \tilde{p}_{\parallel}}_{\ell}(k) = {2\ell + 1 \over 2} \int_{-1}^{1} \grave{P}^{\tilde{\delta}, \tilde{p}_{\parallel}}(\mu, k) {\cal L}_{\ell}(\mu) d\mu .
\end{equation}

The non-linear power spectra (Eqs.\ref{eq:th1},\ref{eq:th2}) are predicated on the global plane parallel assumption -- there is a constant line of sight vector against which the angle $\mu$ between it and Fourier modes can be defined. In contrast, the power spectrum extracted from the data utilises the `local plane parallel approximation', for which ${\cal L}_{\ell}({\bf k} \cdot {\bf \hat{r}}) = {\cal L}_{\ell}({\bf k} \cdot {\bf \hat{r}'})$, where ${\bf \hat{r}}$ and ${\bf \hat{r}'}$ are unit vectors. One can expect discrepancies between theoretical expectation and data due to these two different interpretations of the plane parallel approximation. However, higher order perturbation theory modelling almost exclusively utilises the global plane parallel limit\footnote{The breakdown of the plane parallel limit has been considered in the large scale, Gaussian limit \citep{Castorina:2017inr,Castorina:2019hyr}, although see also \citet{Pardede:2023ddq}.}. Since we expect breakdown of the plane parallel limit to only be relevant on the largest scales, where the statistical error is large, we do not anticipate any significant bias will occur by fitting the global plane parallel approximated power spectra to the data. However, this is a consequence of the modest data volume available; as the data quality improves \citep{DESI:2016fyo,Amendola:2016saw,Dore:2014cca} this issue must be addressed. One possibility is a hybrid splitting of small and large scale modes, as suggested in \citet{Castorina:2017inr}.

The theoretical power spectra are sensitive to a large number of parameters -- both cosmological and those pertaining to galaxy bias. In terms of cosmology, the shape of the galaxy density power spectrum is sensitive to $\Omega_{b}h^{2}$, $\Omega_{c}h^{2}$, $n_{s}$ and the amplitude to $b_{1} \sigma_{8}$, where $b_{1}$ is the linear galaxy bias and $\sigma_{8}$ is the rms fluctuation of the density. In addition, in redshift space both density and momentum power spectra are sensitive to the combination $f \sigma_{8}$. Since we measure the statistics via biased tracers, at third order in perturbation theory we have four additional parameters -- $b_{1}, b_{2}, b_{3, nl}, b_{s}$. The definitions of the higher order bias terms can be found in \citet{McDonald:2009dh,Howlett:2019bky}. To linear order in the perturbations, the linear galaxy bias $b_{1}$ is completely degenerate with $\sigma_{8}$, but this is not the case when including higher order contributions, and we must simultaneously vary $b_{1}$ and $b_{1}\sigma_{8}$. The quantities $P_{ij}$ used in equations (\ref{eq:th1},\ref{eq:th2}) contain a velocity dispersion (labelled $\sigma_{v}$, defined in units of $h^{-1} \, {\rm Mpc}$). We treat $\sigma_{v}$ as an additional free parameter used to model the non-linear Finger-of-God effect. It has been argued elsewhere that multiple velocity dispersion parameters should be used to accurately model the Finger of God effect \citep{Okumura:2015fga}, but for the limited range of scales that we measure, a single parameter is sufficient. 

Some concessions must be made -- we cannot constrain all  parameters from the available data, but fortunately, the power spectra are practically insensitive to a subset of them over the scales probed in this work. For this reason, we fix $\Omega_{b} = 0.048$, $n_{s} = 0.96$, $b_{s} = -4(b_{1} - 1)/7$, and $b_{3,nl} = 32(b_{1}-1)/315$ \citep{Saito:2014qha}. Practically, we have found that the non-linear power spectrum contributions pertaining to $b_{1}$ are the most significant. We fix $\Omega_{b}$ as the baryon fraction has been measured accurately and independently using astrophysical sources and Big Bang Nucleosynthesis (BBN), which is sufficiently robust for our purposes. The primordial tilt $n_{s}$ will affect the shape of the matter power spectrum, but it is measured so accurately by the CMB that we can fix this parameter. We have found that order $\sim 10 \%$ variations of $n_{s}$ will not impact our final constraints, whereas the parameter is constrained to $\sim 0.4\%$ accuracy from the CMB temperature data \citep{Planck:2018vyg}. Although it would be preferable to simultaneously constrain $n_{s}$, $\Omega_{\rm b}$, $\Omega_{m}$, and $\sigma_{8}$, this is simply not feasible with the quality of data currently available. In the literature, multiple parameters pertaining to the shape of the galaxy power spectrum are fixed by convention. This highlights the fact that current velocity catalogues are not competitive cosmological probes compared to the CMB or expansion rate data such as type Ia supernova. However, improvements in both quality and volume of large scale structure data will allow us to further refine our analysis and perform a more complete cosmological parameter estimation in the future \cite{Saulder:2023oqm}.

For similar reasons, the second order bias $b_{2}$ is fixed as $b_{2}=-0.05$. We have tried varying this parameter and found our results to be insensitive to its value over the prior range $-1 \leq b_{2} \leq 1$. Hence we fix $b_{2} = -0.05$, which is the best fit value inferred from the mock galaxies (see Appendix I A). The power spectra are only very weakly sensitive to $b_{s}$ and $b_{3,nl}$ over the scales probed, and our choice of these quantities does not affect our conclusions. We adopt the values quoted in \citep{Saito:2014qha} without any impact to our results. The final list of parameters varied in this work is $\Omega_{m}, b_{1}\sigma_{8}, \sigma_{v}, b_{1}$. The purpose of this paper is to infer a constraint on $b_{1}\sigma_{8}$ and $f\sigma_{8}$, so we treat $f\sigma_{8} = \Omega_{m}^{6/11} (b_{1}\sigma_{8})/b_{1}$ as a derived parameter.

In Figure \ref{fig:theory}, we present the non-linear theoretical galaxy density (left panel) and momentum (right panel) power spectra, using fiducial parameters $\Omega_{m} = 0.3$, $n_{s} = 0.96$, $\Omega_{b} = 0.048$, $\sigma_{8} = 0.8$, $b_{1} = 1.2$, $b_{2} = -0.05$,  $b_{s} = -4(b_{1} - 1)/7$, and $b_{3,nl} = 32(b_{1}-1)/315$ \citep{Saito:2014qha}. We allow the velocity dispersion to take values $\sigma_{v} = 5, 6, 7 h^{-1} {\rm Mpc}$ (solid pink, black, tan lines) . We also present the linear power spectra as black dashed lines. The momentum power spectrum significantly departs from its linear limit even on intermediate scales $k \sim 0.05 \, h \, {\rm Mpc}^{-1}$, and the velocity dispersion $\sigma_{v}$ suppresses $\grave{P}^{\tilde{p}_{\parallel}}$ and $\grave{P}^{\tilde{\delta}}$, with the suppression in $\grave{P}^{\tilde{p}_{\parallel}}$ entering on larger scales than $\grave{P}^{\tilde{\delta}}$.

\begin{figure*}
 \begin{center}
  \includegraphics[width=0.9\textwidth]{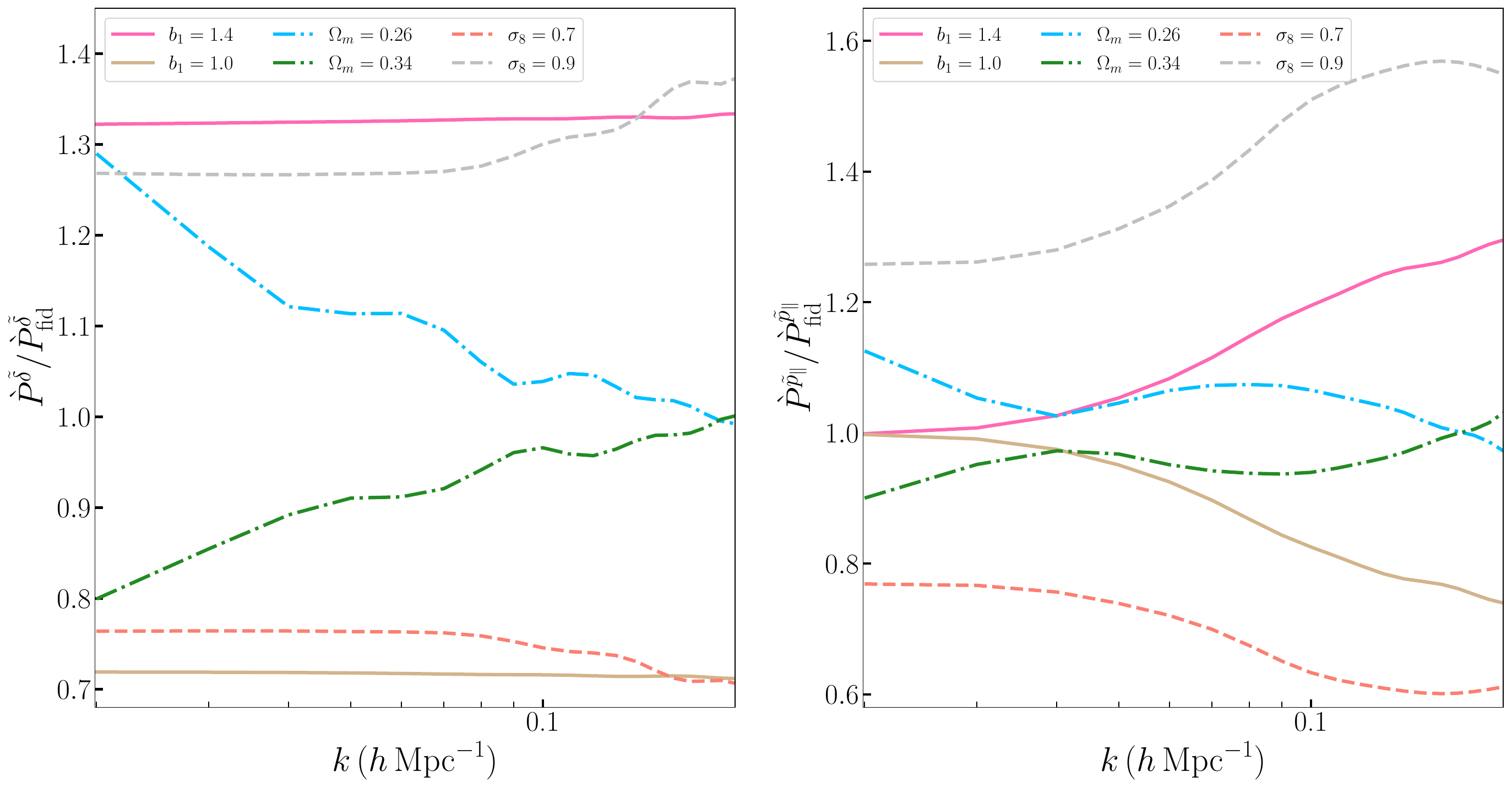}
 \end{center}
 \caption{\label{fig:sensitivity} The ratio $\grave{P}^{\tilde{\delta}}(\Omega_{m},\sigma_{8},b_{1})/\grave{P}^{\tilde{\delta}}_{\rm fid}$ (left panel) and $\grave{P}^{\tilde{p}_{\parallel}}(\Omega_{m},\sigma_{8},b_{1})/\grave{P}^{\tilde{p}_{\parallel}}_{\rm fid}$ (right panel) for different values of the parameters $(\Omega_{m}, \sigma_{8}, b_{1})$, relative to a fiducial parameter set $(\Omega_{m}, \sigma_{8}, b_{1}) = (0.3, 0.8, 1.2)$. 
}
\end{figure*}

In the literature, the quantity $f\sigma_{8}$ is normally treated as a free parameter to be constrained by the data. In this work, we take a different approach by varying the standard cosmological and bias parameters to infer $f \sigma_8$. Our reasoning is that the theoretical ensemble averages to which we compare the data to are entirely derived using General Relativity and the $\Lambda$CDM model, and for this model, $f$ is fixed in terms of other cosmological parameters. Furthermore, beyond linear theory, the redshift-space momentum power spectrum depends on $f$ due to higher order perturbative terms, thus only changing $f\sigma_8$ cannot fully accommodate the dependence of the theory on $f$.
Varying $\Omega_m$ allows us to predict the theoretical templates as accurately as possible. Using our approach, departures from the standard gravitational model would be detectable, via cosmological parameter posteriors that differ significantly from those obtained from other data sets such as the CMB. To map between parameters, we use the approximation $f \simeq \Omega_{m}^{\gamma}$ with $\gamma = 6/11$, which is accurate to sub-percent level for the parameter ranges considered in this work \citep{Linder:2005in}. In non-standard gravity models, the exponent $\gamma$ can acquire both redshift and scale dependence \citep{Appleby:2010dx}.

The premise of utilising the velocity field in cosmological parameter estimation is that the galaxy power spectrum amplitude is sensitive to both $b_{1}\sigma_{8}$ and $f\sigma_{8}$, whereas the amplitude of the velocity  power spectrum only to $f\sigma_{8}$, which breaks the degeneracy. However, the actual picture is complicated by two issues. First, the momentum power spectrum is a linear combination of various two-point functions, the dominant two being $\langle v_{\parallel}({\bf x}) v_{\parallel}(\bf x') \rangle$ and $\langle \delta_{g}({\bf x}) v_{\parallel}({\bf x}) \delta_{g}({\bf x'}) v_{\parallel}({\bf x'}) \rangle$ on large and small scales respectively. Most of the statistical constraining power is at small scales where the $\langle \delta_{g}({\bf x}) v_{\parallel}({\bf x}) \delta_{g}({\bf x'}) v_{\parallel}({\bf x'}) \rangle$ term dominates. As a result, adding the momentum power spectrum to the galaxy density power spectrum does not completely break the degeneracy between cosmological parameters, since $f\sigma_{8}$ and $b_{1}\sigma_{8}$ increase the amplitude of both the momentum power spectrum on small scales and the galaxy density power spectrum on all scales.

Second, we are varying $\Omega_{m}$, and then inferring the growth rate $f \simeq \Omega_{m}^{6/11}$. However, $\Omega_{m}$ also changes the shape of the matter power spectrum, shifting the peak by changing the matter/radiation equality scale. Since we only measure the power spectra over a limited range of scales $0.02 \, h \, {\rm Mpc}^{-1} \leq k \leq 0.2 \, h \, {\rm Mpc}^{-1}$, which does not include the peak, a shift in the peak position can be confused with an increase in amplitude. This introduces an additional source of correlation -- two parameter sets can admit the same values of $b_{1}\sigma_{8}$ and $f\sigma_{8}$ but present seemingly different power spectrum amplitudes over the range of scales probed. This introduces a three-way correlation between $b_{1}\sigma_{8}$, $f\sigma_{8}$ and $b_{1}$.

We present the parameter sensitivity of the theoretical power spectra -- galaxy density (left panel) and momentum field (right panel) -- in Figure \ref{fig:sensitivity}. We take a fiducial set of parameters $(\Omega_{m}, \sigma_{8}, b_{1}) = (0.3, 0.8, 1.2)$, and then vary each separately, fixing $\sigma_{v} = 6 \, (h^{-1} \, {\rm Mpc})$. In the Figure we plot the ratio of power spectra $\grave{P}^{\tilde{\delta}}(\Omega_{m},\sigma_{8},b_{1})/\grave{P}^{\tilde{\delta}}_{\rm fid}$ and $\grave{P}^{\tilde{p}_{\parallel}}(\Omega_{m},\sigma_{8},b_{1})/\grave{P}^{\tilde{p}_{\parallel}}_{\rm fid}$, where the denominator is the power spectrum assuming the fiducial parameter set. We see that the parameters $b_{1}$ and $\sigma_{8}$ do not have a degenerate effect, as $\sigma_{8}$ changes both the overall amplitude and shape of  $\grave{P}^{\tilde{p}_{\parallel}}$ whereas $b_{1}$ only changes the shape. The sensitivity of the power spectra to $\Omega_{m}$ arises from both $f \sim \Omega_{m}^{\gamma}$ and the matter/radiation equality scale. We observe that variation of $\Omega_{m}$ does not correspond only to an amplitude shift for either momentum or galaxy density power spectrum -- the shape is sensitive to this parameter and we are extracting information from both the amplitude and shape of these statistics. 

We note that many of these issues will be ameliorated by reducing the statistical error associated with the large scale modes, by increasing the volume of data. Given the current data limitations, a three-way correlation between $b_{1}$, $f\sigma_{8}$ (or $\Omega_{m}$) and $b_{1}\sigma_{8}$ is unavoidable.

Cosmological parameter dependence enters into our analysis in another way. When measuring the observed power spectra, we bin galaxies into a three-dimensional grid, using redshift and angular coordinates to generate comoving distances. This procedure is sensitive to $\Omega_{m}$, but only very weakly because the sample occupies a low redshift $z \leq 0.055$. Hence when generating the field we fix $\Omega_{m} =0.3$. The parameter $h$ is absorbed into distance units throughout, but we must adopt a value of $h$ when generating the theoretical matter power spectrum. We take $h = h_{\rm pl} =0.674$ \citep{Planck:2018vyg} for our analysis of the SDSS data, and $h = h_{\rm HR4}=0.72$ when applying our methodology to mock data, since this is the value used in the simulation.

\subsection{Mask Convolution}

The mask acts as a multiplicative window function on the field in configuration space, and hence a convolution in Fourier space. The power spectra extracted from the data is implicitly convolved, so to compare the measurement with the theoretical expectation value one must either de-convolve the data and mask, or convolve the theoretical power spectrum with the mask\footnote{In contrast, statistics that measure the one-point information of the field are practically insensitive to the mask \citep{Appleby:2021xoz}.}. We take the latter approach, which is more common in power spectrum analyses. One complication is the fact that the mask will couple the $\ell$ modes of the power spectrum due to the convolution. To proceed, we start with the $\ell$-modes of the unmasked power spectra $P_{\ell}{}^{\tilde{\delta}}$ and $P_{\ell}{}^{\tilde{p}_{\parallel}}$, and perform a Hankel transform to obtain the corresponding real space correlation function $\ell$-modes -- 

\begin{equation}\label{eq:corfn} \zeta_{\ell}{}^{\tilde{\delta}, \tilde{p}_{\parallel}}(r) = {(-1)^{\ell} \over 2\pi^{2}} \int k^{2}dk P_{\ell}{}^{\tilde{\delta}, \tilde{p}_{\parallel}}(k) j_{\ell}(kr) . 
\end{equation} 

\noindent We then take the product of $\zeta_{\ell}{}^{\tilde{\delta}, \tilde{p}_{\parallel}}(r)$ with the mask $\ell$-modes \citep{Wilson:2017} -- 

\begin{eqnarray} {}^{c}\zeta_{0}{}^{\tilde{\delta}, \tilde{p}_{\parallel}} (r) = \zeta_{0}{}^{\tilde{\delta}, \tilde{p}_{\parallel}}Q_{0} + {1 \over 5} \zeta_{2}{}^{\tilde{\delta}, \tilde{p}_{\parallel}}Q_{2} + {1 \over 9} \zeta_{4}{}^{\tilde{\delta}, \tilde{p}_{\parallel}} Q_{4} + ...  
\end{eqnarray} 

\noindent where the ${}^{c}$ superscript denotes that the quantity has been convolved with the mask and $Q_{\ell}(r)$ are the $\ell$-modes of the real space mask correlation function, computed from a random point distribution chosen to match the angular and radial selection of the survey volume. ${}^{c}\zeta_{0}{}^{\tilde{\delta}, \tilde{p}_{\parallel}}$ are the $\ell = 0$ mode correlation functions, corrected by the presence of the mask. Finally, the corresponding monopole power spectrum is inferred by inverse Hankel transforming 

\begin{equation} {}^{c}P_{0}{}^{\tilde{\delta}, \tilde{p}_{\parallel}}(k) =  4\pi \int r^{2} dr \, {}^{c}\zeta_{0}{}^{\tilde{\delta}, \tilde{p}_{\parallel}}(r) j_{0}(kr) .
\end{equation} 

\begin{figure}
 \begin{center}
  \includegraphics[width=0.45\textwidth]{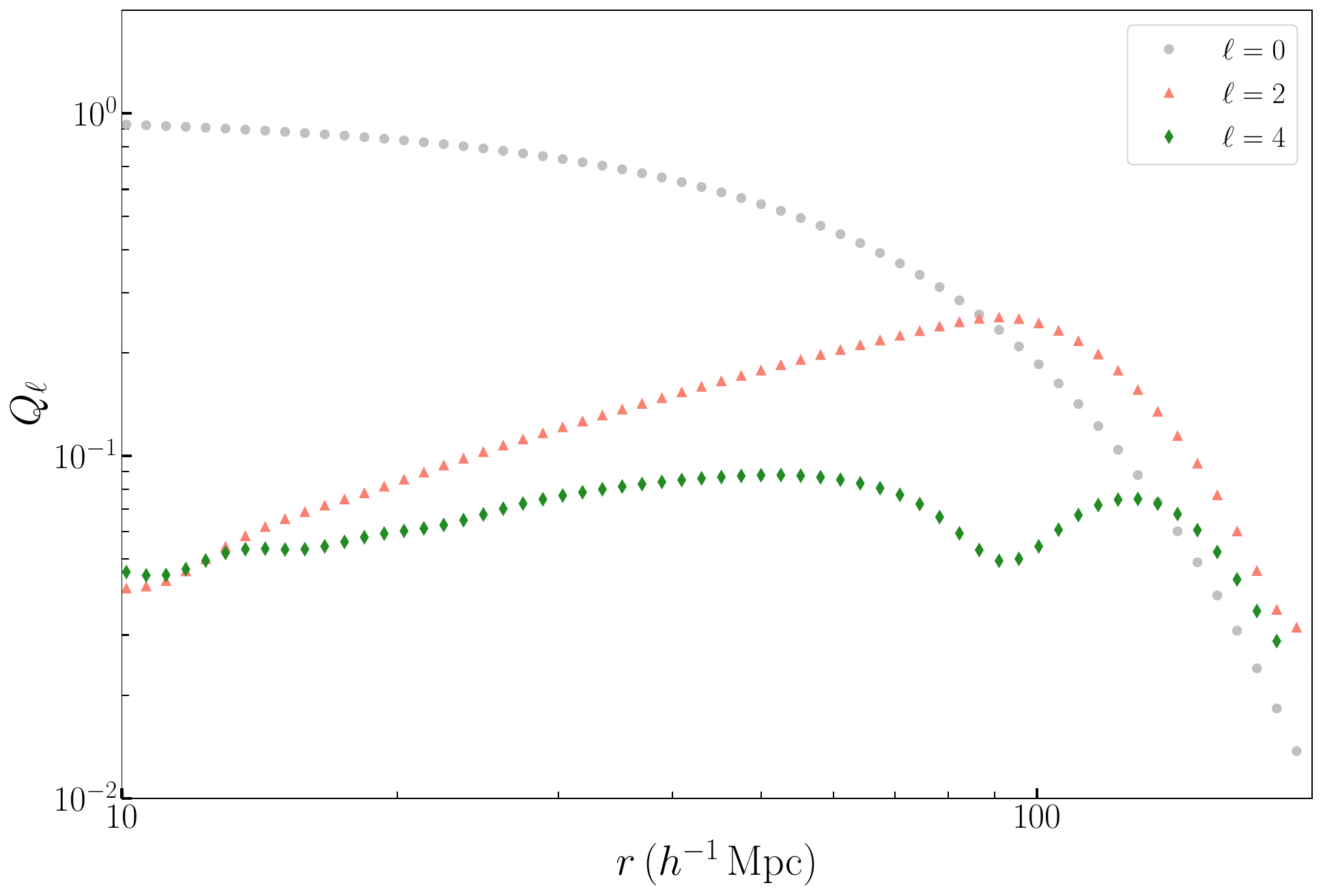}
 \end{center}
 \caption{\label{fig:Q} The $\ell=0,2,4$ modes of the real space mask correlation function, obtained by pair counting a random point distribution that matches the radial and angular selection functions of the data. The $\ell=2$ mode is negative, so we plot $|Q_{2}|$.
}
\end{figure}

\noindent We only extract the $\ell=0$ mode of the power spectrum from the data, since the data volume is not sufficient to obtain an accurate measurement of the higher multipoles. The $\ell$-modes of the mask, $Q_{\ell}(r)$, are obtained by generating a random point distribution of $N=10^{6}$ points which encompass the radial and angular domain of the data, then binning pairs of points as a function of radial separation and angle $\mu$ and finally integrating out the $\mu$ dependence using the orthogonality of the Legendre polynomials. The function $Q_{\ell}$ is normalised such that $Q_{0}(0) = 1$. We present the $Q_{0,2,4}(r)$ modes of the mask correlation function in Figure \ref{fig:Q}.

To perform the masking procedure, in (\ref{eq:corfn}) we use the power spectra $P_{\ell}{}^{\tilde{\delta}}$ and $P_{\ell}{}^{\tilde{p}_{\parallel}}$ measured from a mock galaxy catalog in a $z=0$ snapshot box, using a simulation described in the following section. After performing the mask convolution and arriving at ${}^{c}P_{0}{}^{\tilde{\delta}, \tilde{p}_{\parallel}}(k)$, we define the ratios $r^{\tilde{\delta}} = {}^{c}P_{0}{}^{\tilde{\delta}}/P_{0}{}^{\tilde{\delta}}$ and $r^{\tilde{p}_{\parallel}} = {}^{c}P_{0}{}^{\tilde{p}_{\parallel}}/P_{0}{}^{\tilde{p}_{\parallel}}$. To adjust the theoretical power spectra (\ref{eq:th1},\ref{eq:th2}) to account for the effect of the mask, we multiply them by these pre-computed ratios. We adopt this approach, rather than applying the Hankel transforms directly to the theoretical expectation values (\ref{eq:th1},\ref{eq:th2}), because the perturbative expansion breaks down on small scales and this could generate artificial numerical artifacts when performing the $k$-space integrals. Practically, we find no significant change to the convolution when we include the $Q_{\ell=2}$, $Q_{\ell=4}$ modes of the mask, compared to just using the $Q_{\ell=0}$ component in the convolution. We include them for completeness only.

Since we only extract the $\ell=0$ mode from the data, to simplify our notation in what follows we drop the $\ell = 0$ subscript and simply denote the measured(theoretical) monopole power spectra as $P^{\tilde{\delta}, \tilde{p}_{\parallel}}$($\grave{P}^{\tilde{\delta}, \tilde{p}_{\parallel}}$), and the corresponding `convolved' theoretical power spectra are ${}^{c}\grave{P}^{\tilde{\delta}} = r^{\tilde{\delta}} \grave{P}^{\tilde{\delta}}$ and ${}^{c}\grave{P}^{\tilde{p}_{\parallel}} = r^{\tilde{p}_{\parallel}} \grave{P}^{\tilde{p}_{\parallel}}$.

\section{Mock Data and Covariance Matrices} 
\label{sec:mocks}

We initially extract the galaxy density and momentum power spectra from the Horizon Run 4 simulation $z=0$ snapshot box \citep[HR4;][]{Kim:2015yma}, and then use this mock data to generate a covariance matrix for the SDSS data. We briefly describe the simulation used and present power spectrum and covariance matrix measurements. 

HR4 is a cosmological scale cold dark matter simulation in which $6300^{3}$ particles were gravitationally evolved in a $V = (3150 \, h^{-1} \, {\rm Mpc})^{3}$ box using a modified version of GOTPM code\footnote{The original GOTPM code is introduced in \cite{DUBINSKI2004111}. A review of the modifications introduced in the Horizon Run project can be found at \href{https://astro.kias.re.kr/~kjhan/GOTPM/index.html}{https://astro.kias.re.kr/\textasciitilde kjhan/GOTPM/index.html}}. Details of the simulation can be found in \citet{Kim:2015yma}. 

Dark Matter (DM) halos are constructed using the friends-of-friends (FoF) algorithm with a linking length $0.2$ times the mean DM particle separation for each snapshot. Only those with more than $30$ DM particles are identified as halos, making the minimum halo mass $2.7 \times 10^{11} \, h^{-1} \, M_{\odot}$. Then, the merger tree of HR4 DM halos is constructed for $75$ snapshots between $z = 12$ and $0$ with a timestep of $\sim 0.1 {\rm Gyr}$. During the construction of the merger tree, the most-bound halo particle (MBP) of each halo is traced, even when its hosting halo is merged into a more massive halo. 

The HR4 mock galaxy catalog is constructed by following \citet{Hong:2016hsd} using an MBP-galaxy correspondence method. If an MBP is from an isolated halo, its position and velocity are regarded as those of the central galaxy of the given halo. On the other hand, if an MBP is from a halo already merged into a more massive halo, it is regarded as a candidate for a satellite galaxy. Then, the survival time of the satellite candidate after merger event $t_{\rm merge}$ is calculated using a modified model of \citet{Jiang:2007xd}, 

\begin{equation} {t_{\rm merge} \over t_{\rm dyn}} = {\left(0.94\epsilon^{0.6} + 0.6 \right)/0.86 \over \ln \left[ 1 + (M_{\rm host}/M_{\rm sat})\right]} \left({M_{\rm host} \over M_{\rm sat}} \right)^{1.5} , \end{equation} 

\noindent where $t_{\rm dyn}$ is the orbital period of a virialized object; $M_{\rm host}$ and $M_{\rm sat}$ are the masses of host and satellite halos, and epsilon is the circularity of the satellite's orbit. Only the satellite candidates whose merger events occurred no earlier than $t_{\rm merge}$ are successfully identified as satellite galaxies. 

Typically, dark matter only simulations underestimate the population of satellites in dark matter halos and the two-point correlation function imperfectly matches observations as the one-halo term is underestimated. In our scheme the timescale for satellites to merge
with the central is taken into account by tracking the most-bound halo particles at the time of
their dark matter halo entrance regardless of survival of the dark matter
subhalo. By adjusting the timescale we can match the correlation function of the simulated
galaxies with that of the SDSS Main Galaxy Sample \citep{SDSS:2001fzq}. We direct the reader to Figure 1 of \citet{Hong:2016hsd} for a complete description. With our choice of algorithm, we do not have the freedom to vary the small-scale distribution of satellite galaxies; this is fixed by following the MBP dynamics and selecting the timescale to match the two-point clustering of the data.

We select mock galaxies by applying two independent mass cuts -- one to central galaxies and the second to satellites, such that the total number of galaxies in the snapshot box is $N=1.66 \times 10^{8}$ and the satellite fraction is $f_{\rm sat} = 0.4$. These values were chosen to match the number density of the SDSS FP catalog $\bar{n} = 5.3 \times 10^{-3} \, h^{3} \, {\rm Mpc}^{-3}$ and roughly match the expected satellite fraction of ETGs \citep{Mandelbaum:2005nx}. The central/satellite mass cuts for the HR4 mock galaxies are $M_{\rm cen} = 9.3\times 10^{11} M_\odot/h$ and $M_{\rm sat} = 6.3\times10^{11} M_\odot/h$, respectively. 

We generate a regular grid of resolution $\Delta = 3150/768 = 4.1 \, h^{-1} \, {\rm Mpc}$ in each dimension, covering the snapshot box and assign each galaxy to a pixel according to a
CIC scheme. For the scales being considered in this work -- $0.02 \, h \, {\rm Mpc}^{-1} \leq k \leq 0.2 \, h \, {\rm Mpc}^{-1}$, we have checked that there are no resolution effects due to our choice of binning. 

We generate real and redshift-space fields from the data. In real space, galaxy positions are given by their comoving position within the box. To generate a redshift space field, the position of each galaxy is perturbed according to equation (\ref{eq:rsp}), where we take ${\bf \hat{e}}_{\parallel} = {\bf \hat{e}}_{z}$ so that the line of sight vector is aligned with the $z$-Cartesian component of the box. Because we are using the redshift zero snapshot box and absorbing $h$ into distance units, we take $(aH)^{-1} = 10^{-2}h^{-1}$ as the numerical factor between velocity and distance. We denote the real and redshift space galaxy density fields as $\delta$ and $\tilde{\delta}$ respectively, where $\delta_{ijk} = (n_{ijk} - \bar{n})/\bar{n}$, $\bar{n}$ is the average galaxy per pixel, and $n_{ijk}$ is the number of particles in the $(i, j, k)$ pixel in a cubic lattice.  

We also generate momentum fields by assigning galaxy velocities to the same grid as the number density. The momentum field studied in this section is given by $p^{(g)}_{ijk} = v^{( g)}_{ijk}/\bar{n}$, where $v^{(g)}_{ijk}$ is the sum of all galaxy velocities assigned to the $(i, j, k)$ pixel. We can extract $p_{ijk}^{(g)}$ from the snapshot using the real and redshift space positions of the galaxies, which we denote with/without a tilde respectively. The galaxy momentum field in redshift space -- $\tilde{p}^{(g)}$ -- is the quantity that we will measure from the data.

\begin{figure}
    \centering
    \includegraphics[width=0.45\textwidth]{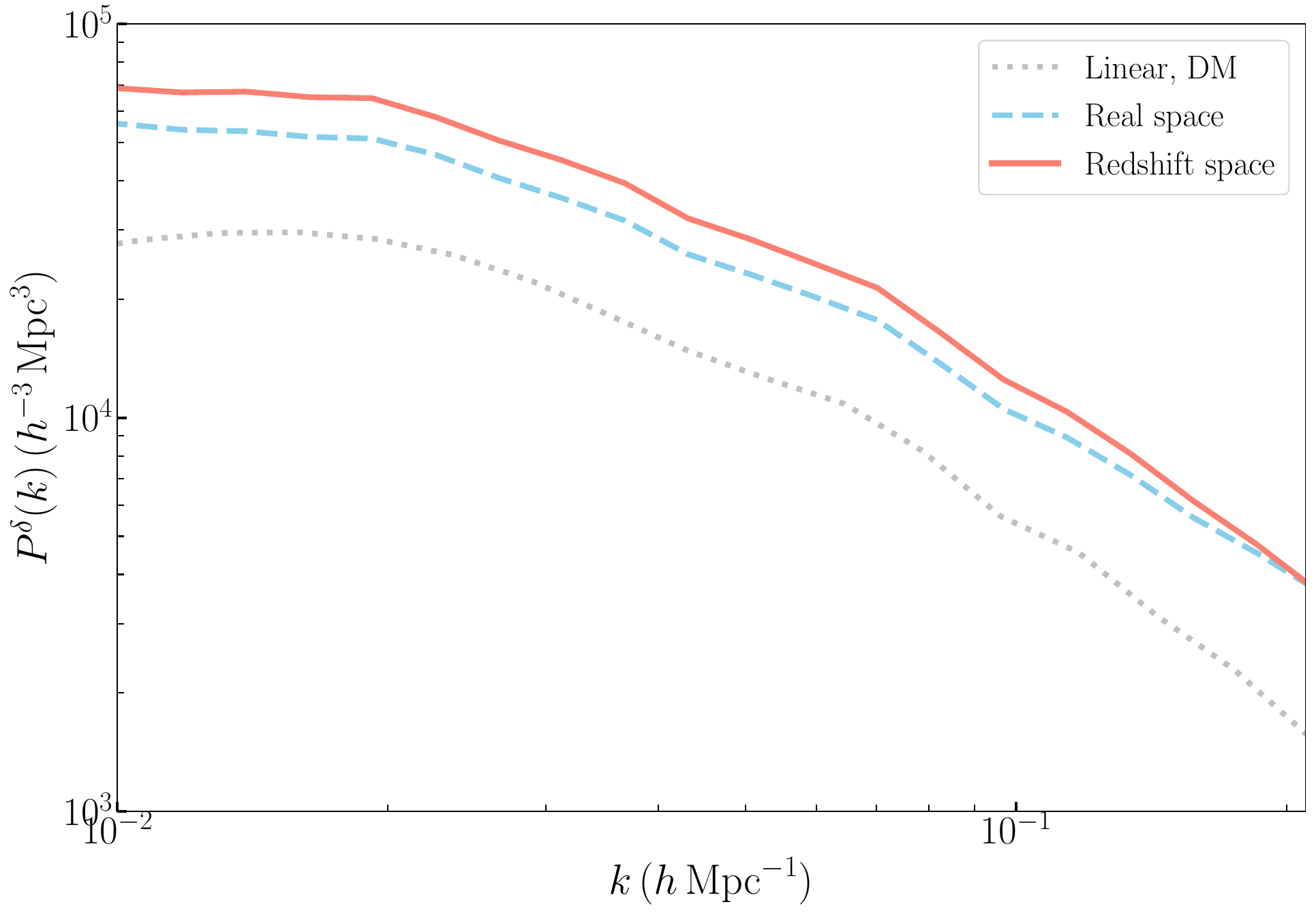}
    \includegraphics[width=0.45\textwidth]{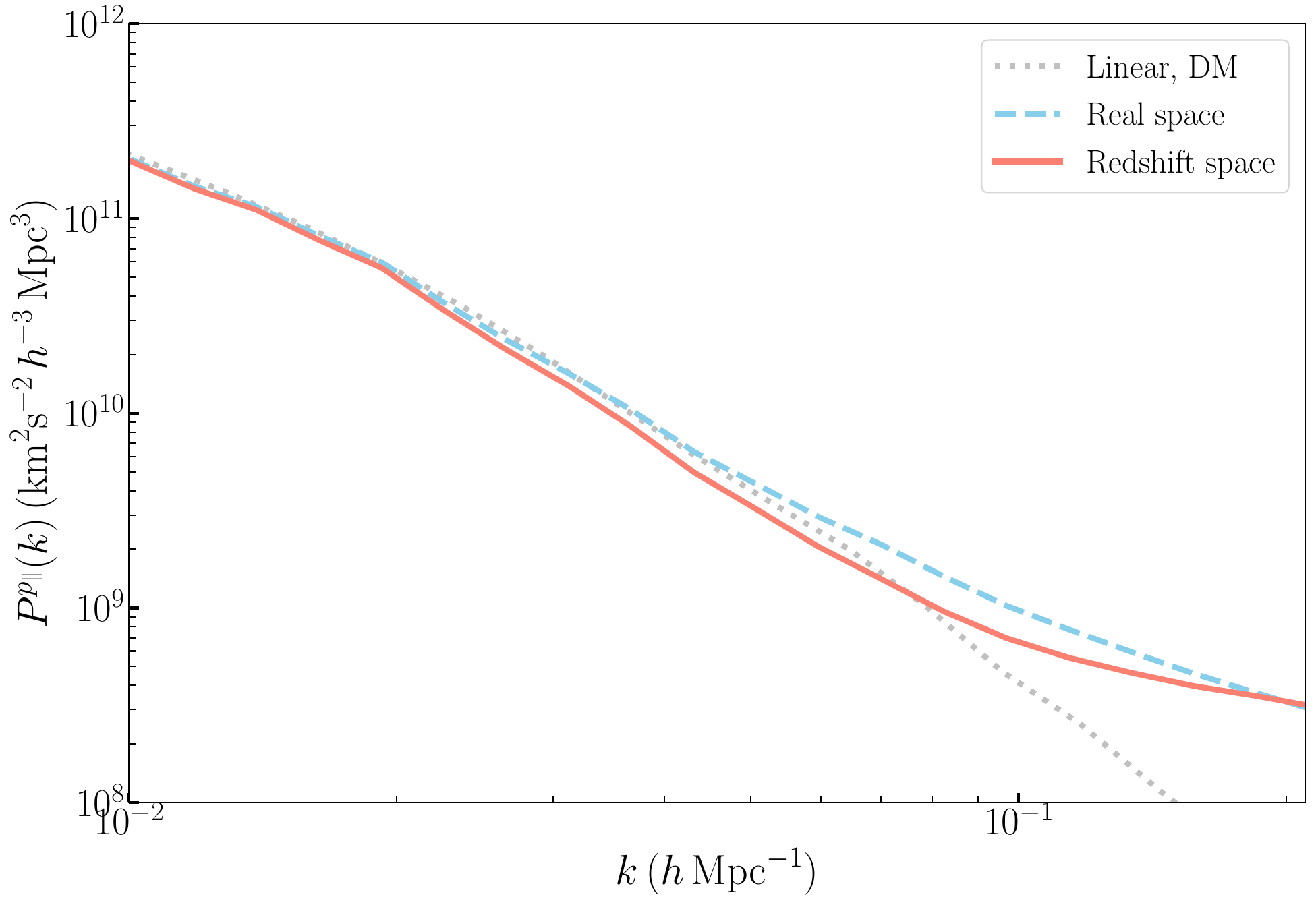}
    \caption{[Left panel] The galaxy density power spectrum extracted from the HR4 $z=0$ snapshot box in real/redshift space (blue dashed/red solid). The grey dotted line is the linear dark matter (DM) power spectrum. [Right panel] The momentum power spectrum, with the same colour scheme as in the left panel. }
    \label{fig:1}
\end{figure}

In Figure \ref{fig:1}, we present the galaxy density power spectra (left panel) and momentum power spectra (right panel). In both panels, the dotted grey curve is the linear, dark matter density and momentum expectation value, and the blue dashed, red solid lines are the galaxy density/momentum power spectra in real, redshift space respectively. 

The matter power spectrum presents familiar results -- the mock galaxies have a linear bias $b_{1} \simeq 1.3$ and the effect of redshift space distortion amplifies the power spectrum amplitude on large scales by an additional factor of $1 + 2\beta/3 + \beta^{2}/5$, with $\beta = f/b_{1}$. On small scales, starting at $k \gtrsim 0.1 \, h \, {\rm Mpc}^{-1}$, the redshift space power spectrum exhibits a suppression in power due to galaxy stochastic velocities within their host halos (cf. red curve, left panel). 

The momentum field presents very different behaviour. On the largest scales measurable by the simulation $k \sim 10^{-2} \, h \, {\rm Mpc}^{-1}$, the real and redshift space momentum power spectra agree with the linear theory expectation value with no velocity bias \citep{Zheng:2014vla,Chen:2018ntg}. However, on large scales $0.05 \, h \, {\rm Mpc}^{-1} < k < 0.1 \, h \, {\rm Mpc}^{-1}$ the galaxy redshift space power spectra departs from the real space measurement. It is clear that the shape of the momentum power spectrum is significantly modified by nonlinear effects. To linear order in the fields, the real and redshift space power spectra should be indistinguishable and so the deviation between the red and blue curves in the right panel are due to a combination of higher order perturbative, and fully non-perturbative, non-linear effects. In Appendix I B we consider the impact of non-linear velocity contributions on the momentum power spectrum further.

Next, we generate a covariance matrix for the SDSS data, using a set of $N_{\rm real} = 512$ non-overlapping mock SDSS catalogs from the snapshot box, placing mock observers in the box then applying the SDSS angular footprint and radial selection function relative to the observer position. When generating the mocks we take radial velocities relative to the observer and galaxy positions are corrected according to equation (\ref{eq:rsp}), where we take ${\bf \hat{e}}_{\parallel} = {\bf \hat{e}}_{r}$ which is the radial unit vector pointing between observer and galaxy. The observers are placed in the rest frame of the snapshot box and the radial velocities of the galaxies are used to generate the momentum field.  

For each mock catalog, we measure the galaxy density and momentum power spectra using the same methodology as for the SDSS data, then define a set of covariance matrices as 

\begin{equation} \Sigma_{ij}^{(m,n)} = {1 \over N_{\rm real}-1} \sum_{q=1}^{N_{\rm real}}  \left( y_{q, i}^{(m)} - \bar{y}_{i}^{(m)}\right) \left( y_{q, j}^{(n)} - \bar{y}_{j}^{(n)}\right) 
\end{equation}

\begin{figure*}
    \centering
    \includegraphics[width=0.98\textwidth]{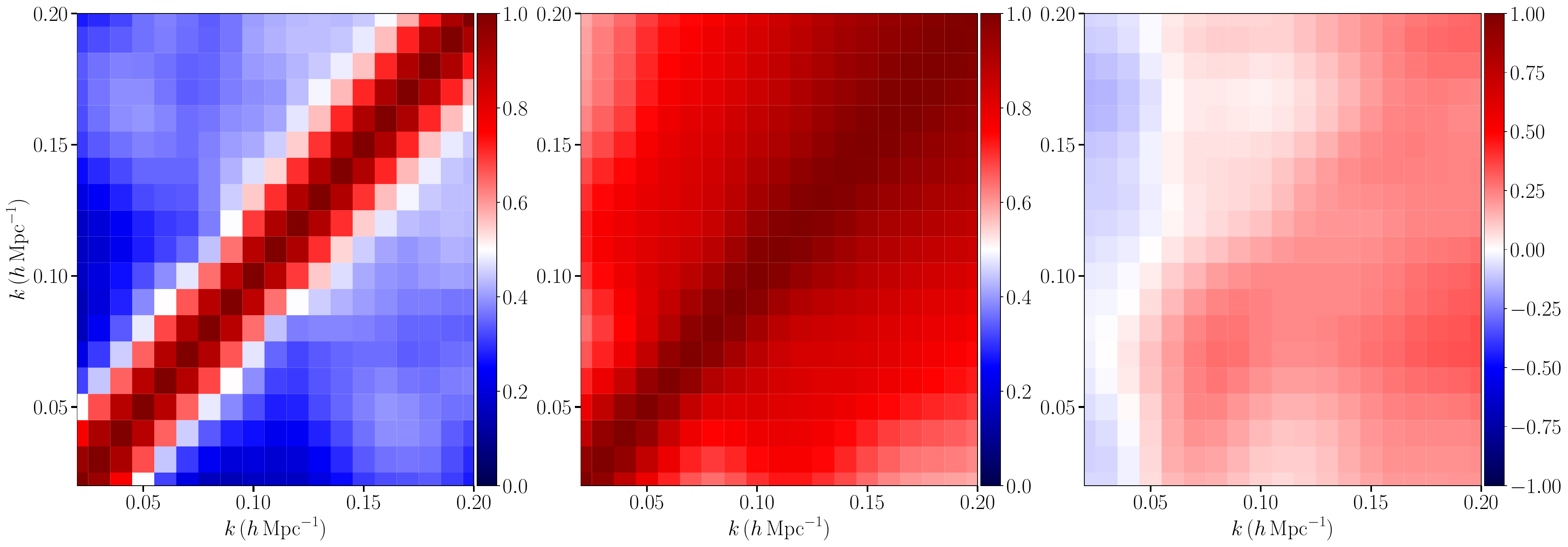}  
    \caption{ Correlation matrices for $(\tilde{\delta}, \tilde{\delta})$ (left panel), $(\tilde{p}_{\parallel},\tilde{p}_{\parallel})$ (middle panel) and $(\tilde{\delta}, \tilde{p}_{\parallel})$ (right panel).}
    \label{fig:correlation}
\end{figure*}

\noindent where $1 \leq i,j \leq N_{k}$ denote the Fourier mode bin $k_{i}, k_{j}$, $(m,n) = (\tilde{\delta}, \tilde{\delta}), (\tilde{\delta}, \tilde{p}_{\parallel}), (\tilde{p}_{\parallel}, \tilde{\delta}), (\tilde{p}_{\parallel}, \tilde{p}_{\parallel})$, $y_{q, i}^{(m)} = \ln[P_{q, i}^{(m)}]$  and $\bar{y}_{i}^{(m)}$ is the sample mean of $y_{q, i}^{(m)}$ in the $i^{\rm th}$ Fourier bin. We use $N_{k} = 19$ Fourier bins, linearly equi-spaced over the range $k=[0.02, 0.2] \, h \, {\rm Mpc}^{-1}$. Explicitly, $P^{(m)}_{q, i}$ is the measured value of the $m=\tilde{\delta}, \tilde{p}_{\parallel}$ power spectrum from the $q^{\rm th}$ realisation in the $i^{th}$ Fourier mode bin. We use $y_{q, i}^{(m)}$ -- the logarithm of the power spectra -- as the observable, as the mocks indicate that the $P_{q, i}^{(m)}$ measurements are not Gaussian distributed within the $k_{i}$ bins. The logarithm is sufficiently Gaussianized for our purposes. In \citet{Qin:2019axr} a different transformation was used -- the Box-Cox transform -- to the same effect. 

Finally, the total covariance matrix is the $2N_{k} \times 2N_{k}$ symmetric, block matrix constructed from these three matrices;

\begin{equation}\label{eq:lam} \Lambda \equiv \begin{pmatrix}
\Sigma^{(\tilde{\delta},\tilde{\delta})} & \Sigma^{(\tilde{\delta},\tilde{p}_{\parallel})} \\
\Sigma^{(\tilde{p}_{\parallel}, \tilde{\delta})} & \Sigma^{(\tilde{p}_{\parallel}, \tilde{p}_{\parallel})}
\end{pmatrix} ,
\end{equation}

\noindent where $\Sigma^{(\tilde{p}_{\parallel}, \tilde{\delta})} = \Sigma^{(\tilde{\delta},\tilde{p}_{\parallel})}$. The inverse of the covariance matrix $\Lambda^{-1}$ is used for parameter estimation. 

In Figure \ref{fig:correlation}, we present the normalised correlation matrices 

\begin{equation} \Theta^{(m,n)}_{ij} = {\Sigma_{ij}^{(m,n)} \over \sigma_{i}^{(m)}\sigma_{j}^{(n)}} , \end{equation}

\noindent where $\sigma_{i}^{(m)}$ is the standard deviation of the mocks in the $i^{th}$ Fourier bin (we are not using the Einstein summation convention). The left/middle/right panels are the $(m,n) = (\tilde{\delta},\tilde{\delta}), (\tilde{p}_{\parallel},\tilde{p}_{\parallel}), (\tilde{\delta}, \tilde{p}_{\parallel})$ matrices, respectively. We note the strong correlation between all Fourier bins of the momentum power spectrum (middle panel). The third panel shows an important property of the momentum field -- it is positively correlated with the galaxy density field \citep{Park:2000rc}. By constructing statistics from the ratio of these fields, one may expect a reduction in cosmic variance and improved constraints on cosmological parameters. This possibility will be considered in the future. 

\begin{figure}
    \centering
    \includegraphics[width=0.45\textwidth]{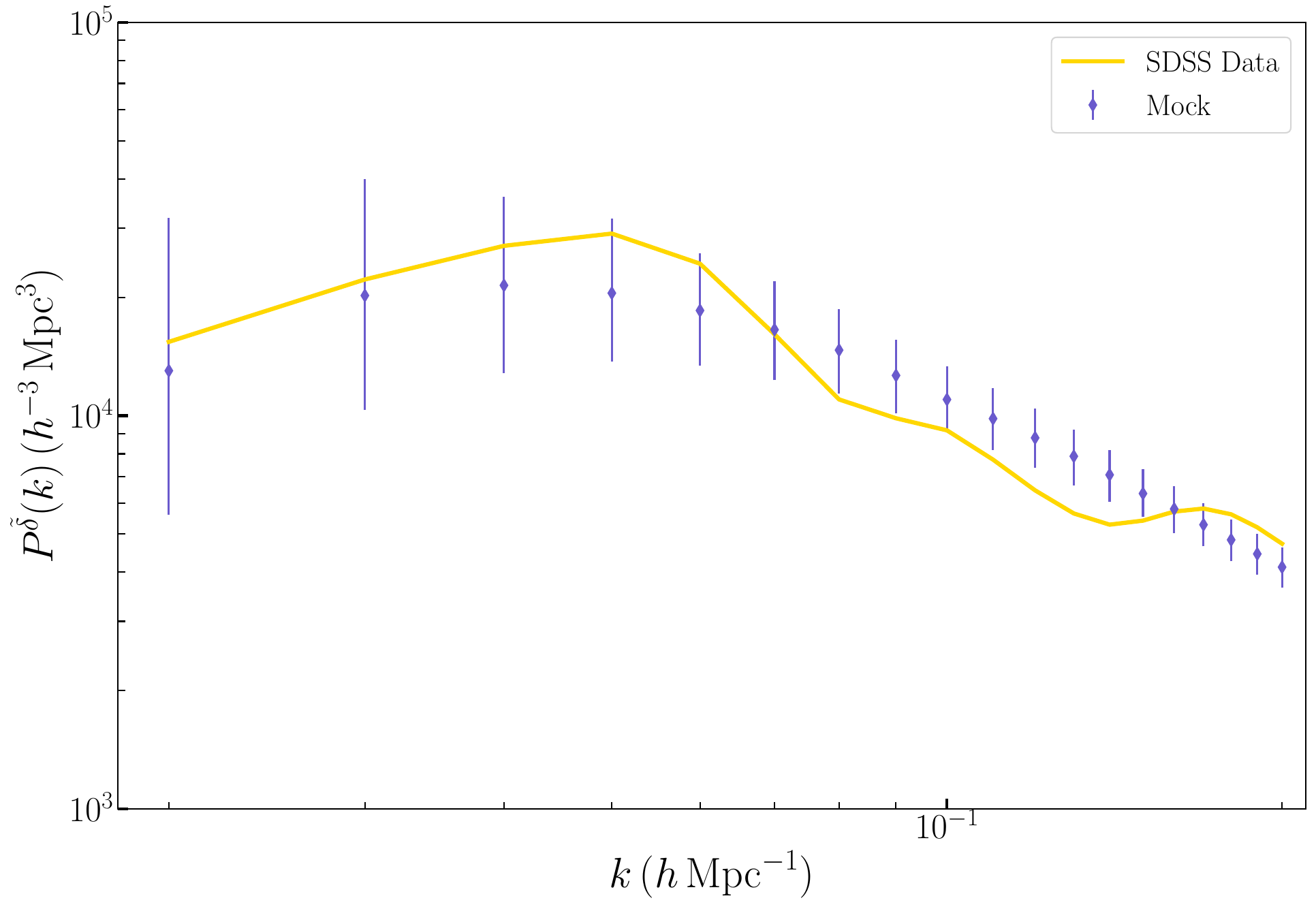}
    \includegraphics[width=0.45\textwidth]{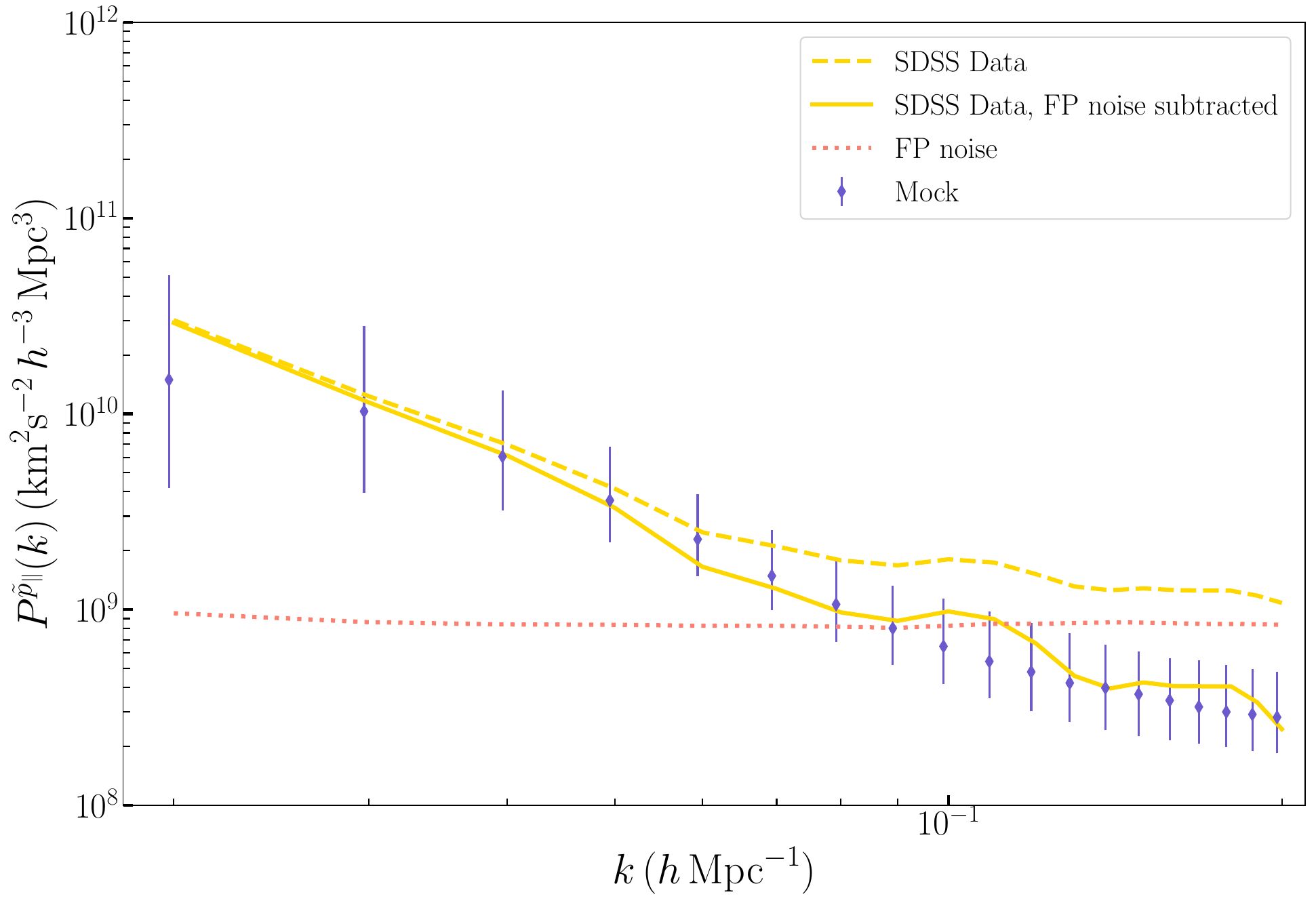}\\
    \caption{[Left panel] Galaxy density power spectrum extracted from the SDSS FP data (solid gold line) and the median and $68\%$ limits from $N=512$ mock catalogs (blue points/error bars). [Right panel] The momentum power spectrum extracted from the SDSS FP data (gold dashed line), the contribution from the FP velocity uncertainty (pink dashed line) and the FP noise-subtracted momentum power spectrum (solid gold line). The blue points/error bars are the median and $68\%$ range inferred from the mock catalogs. }
    \label{fig:data}
\end{figure}

\section{Results}
\label{sec:results}

We now present measurements of the galaxy density and momentum power spectra from the SDSS FP data, and then fit the standard model to the data to infer parameter constraints. 

\subsection{Power Spectra Measurements} 

In Figure \ref{fig:data}, we present the galaxy density (left panel) and momentum (right panel) power spectra extracted from the SDSS velocity catalog (gold lines). We also plot the median and sample $68\%$ range of the $N_{\rm real} = 512$ mock catalogs as blue points/error bars. The power spectrum of the FP data (cf. gold solid line, left panel) exhibits large fluctuations as a result of the modest volume of data -- $0.025 \leq z \leq 0.055$. In particular, on intermediate scales $0.05 \, h \, {\rm Mpc}^{-1} \leq k \leq 0.15 \, h \, {\rm Mpc}^{-1}$ there is less power in the SDSS galaxies relative to the mocks. This is likely a statistical fluctuation.

In the right panel, the blue points are again the median and $68\%$ range of the mocks and the dashed gold dashed line is momentum power spectrum extracted from the SDSS data. We observe that the SDSS power spectrum is dominated on small scales by the statistical uncertainty associated with the FP velocity measurements. This contribution is larger than the cosmological signal on scales $k \geq 0.1 \, h {\rm Mpc}^{-1}$. To eliminate this contribution, we treat this velocity component as Gaussian noise, uncorrelated with the cosmological signal. Then, from the SDSS FP catalog we generate realisations by drawing a random velocity for each galaxy, using the velocity uncertainty as the width of a Gaussian distribution, then reconstructing the momentum power spectrum of these random velocities. We denote the resulting power spectrum as $P_{\rm fpn}(k_{i})$ (FP noise). We repeat this process $N=1000$ times, and construct the average $\bar{P}_{\rm fpn}(k_{i})$ and standard deviation $\Delta P_{\rm fpn}(k_{i})$. The pink dotted line is the resulting FP noise power spectrum $\bar{P}_{\rm fpn}$, and the solid gold line is the SDSS momentum power spectrum with this noise component subtracted. 
To account for the uncertainty in the FP noise removal, the quantity $\Delta P_{\rm fpn}$ is added to the diagonal elements of the covariance matrix $\Sigma^{(\tilde{p}_{\parallel},\tilde{p}_{\parallel})}_{ij}$ -- 

\begin{equation} \Sigma^{(\tilde{p}_{\parallel},\tilde{p}_{\parallel})}_{ij} \to \Sigma^{(\tilde{p}_{\parallel},\tilde{p}_{\parallel})}_{ij} + \left({\Delta P_{\rm fpn}(k_{i}) \over  \bar{P}_{\rm fpn}(k_{i}) } \right)^{2} \delta_{ij} .
\end{equation} 

\noindent Since we assume the FP uncertainty is uncorrelated with the actual velocity, has mean zero and is Gaussian distributed, this component will not contribute to $\Sigma^{(\tilde{\delta},\tilde{\delta})}$ or $\Sigma^{(\tilde{p}_{\parallel},\tilde{\delta})}$. The dimensionless quantity $\Delta P_{\rm fpn}(k_{i}) / \bar{P}_{\rm fpn}(k_{i})$ is of order $\sim {\cal O}(0.1)$ at $k = 0.2 \, h \, {\rm Mpc}^{-1}$.

To confirm our assumption of Gaussianity, in Figure \ref{fig:histograms} of Appendix C we bin the SDSS galaxies into nine comoving distance bins and generate histograms of the velocity uncertainties. The yellow bars are the entire sample, and the dark bars correspond to the galaxies used in our analysis, after a $\Delta v$ cut has been applied. We see that the cut acts to Gaussianize the velocity uncertainty, allowing us to treat this component as Gaussian white noise. 

We take the measured galaxy density and momentum power spectra (cf. solid gold lines, Figure \ref{fig:data}), and fit the mask-convolved theoretical model described in Section \ref{sec:theory}, arriving at constraints on a set of cosmological and galaxy bias parameters. We also apply our analysis pipeline to the mean of the mock realisations to confirm that we generate unbiased parameter constraints.

\begin{table}[tb]
\begin{center}
 \begin{tabular}{||c c c c c ||}
 \hline 
 Parameter \, &  $\Omega_{m}$ \, & $b_{1}\sigma_{8}$ \, &  $b_{1}$ \, &  $\sigma_{v}$ \,  \\ [0.5ex] 
 
 Range \, &  [0.01, 1] \, & [0.2, 3] \, &  [0.2, 10] \, &  [4, 30] \,   \\
  \hline  
\end{tabular}
\caption{\label{tab:1} The parameters varied in this work, and the prior range imposed. The quantity $\sigma_{v}$ has units $h^{-1} \, {\rm Mpc}$. }
\end{center} 
\end{table}

\subsection{Parameter Estimation} 
\label{sec:parm_est}

We first use the mean of the $N=512$ mock realisations to test our analysis pipeline. We define the data vector as $\vec{d} = (y^{\tilde{\delta}}_{i} - \ln \, [{}^{c}\grave{P}^{\tilde{\delta}}_{i}], \,  y^{\tilde{p}_{\parallel}}_{j} - \ln \,[{}^{c}\grave{P}^{\tilde{p}_{\parallel}}_{j}])$, where $y^{\tilde{\delta}}_{i} = \ln \, P^{\tilde{\delta}}_{i}$, $y^{\tilde{p}_{\parallel}}_{j} = \ln \, P^{\tilde{p}_{\parallel}}_{j}$ are the logarithm of the mean of the measured mock galaxy density and momentum power spectra in the $1 \leq i, j \leq N_{\rm k}$ Fourier bins, hence $\vec{d}$ has dimension $2N_{\rm k}$. $\grave{P}^{\tilde{\delta}}_{i}, \grave{P}^{\tilde{p}_{\parallel}}_{j}$ are the theoretical expectation values derived in Section \ref{sec:theory}, and they have been multiplied by the pre-computed effect of the mask convolution; the ratios $r^{\tilde{\delta}}$ and $r^{\tilde{p}_{\parallel}}$ respectively. We minimize the Gaussian likelihood ${\cal L} \propto e^{-\chi^{2}/2}$, with 

\begin{equation} \chi^{2} = \vec{d}^{\rm T} \Lambda^{-1} \vec{d} . \end{equation} 

\noindent We vary the parameter set $(\Omega_{m}, b_{1}\sigma_{8}, b_{1},  \sigma_{v})$, then subsequently infer the derived parameter $f\sigma_{8} \simeq \Omega_{m}^{6/11} (b_{1}\sigma_{8})/b_{1}$. The matter density $\Omega_{m}$ enters the analysis in two ways -- changing the amplitude of the power spectra via $f\sigma_{8}$ and modifying the shape of the linear matter power spectrum via the equality scale $k_{\rm eq}$. 

The prior ranges for our parameters are provided in Table \ref{tab:1}. Reasonable variation of these ranges does not significantly modify our results. We exclude $\sigma_{v} = 0$ from consideration, since this quantity describes the Finger of God effect and must be non-zero. In addition, we apply a weak and physically motivated prior on the galaxy bias, such that $b_{1} = 1_{-0.15}^{+\infty}$. With this choice, we penalise values of the bias $b_{1} < 1$ with an additional Gaussian contribution to the $\chi^{2}$, of width $\sigma_{b_{1}} = 0.15$. We do so because the SDSS galaxies are known to have bias $b_{1} \geq 1$, and the Gaussian width $\sigma_{b_{1}} = 0.15$ is selected as this is the expected statistical constraint on the bias that can be obtained from the volume being probed in this work. In contrast, all values of $b_{1}$ larger than unity are selected with uniform prior, indicated with the $+\infty$ upper bound.

We apply the same methodology to the SDSS extracted power spectra, taking the data vector $\vec{d} = (y^{\tilde{\delta}}_{i} - \ln \, [{}^{c}\grave{P}^{\tilde{\delta}}_{i}], \,  y^{\tilde{p}_{\parallel}}_{j} - \ln \,[{}^{c}\grave{P}^{\tilde{p}_{\parallel}}_{j}])$ with $y^{\tilde{\delta}}_{i} = \ln \, P^{\tilde{\delta}}_{i}$, $y^{\tilde{p}_{\parallel}}_{j} = \ln \, [P^{\tilde{p}_{\parallel}}_{j}-\bar{P}_{{\rm fpn}, j}]$, and $P^{\tilde{\delta}}_{i}$, $P^{\tilde{p}_{\parallel}}_{j}$ are the measured power spectra from the SDSS catalog. We use the same parameter set and ranges as for the mocks shown in Table \ref{tab:1}, covariance matrix $\Lambda$ to generate $\chi^{2}$ for the likelihood and the prior $b_{1} = 1_{-0.15}^{+\infty}$. When generating the theoretical linear matter power spectrum, used in the theoretical power spectra $\grave{P}^{\tilde{\delta}}$ and $\grave{P}^{\tilde{p}_{\parallel}}$, we adopt a value of $h = h_{\rm HR4} = 0.72$ for the mocks and $h = h_{\rm pl} = 0.67$ for the data.

\begin{figure*}
    \centering
    \includegraphics[width=0.45\textwidth]{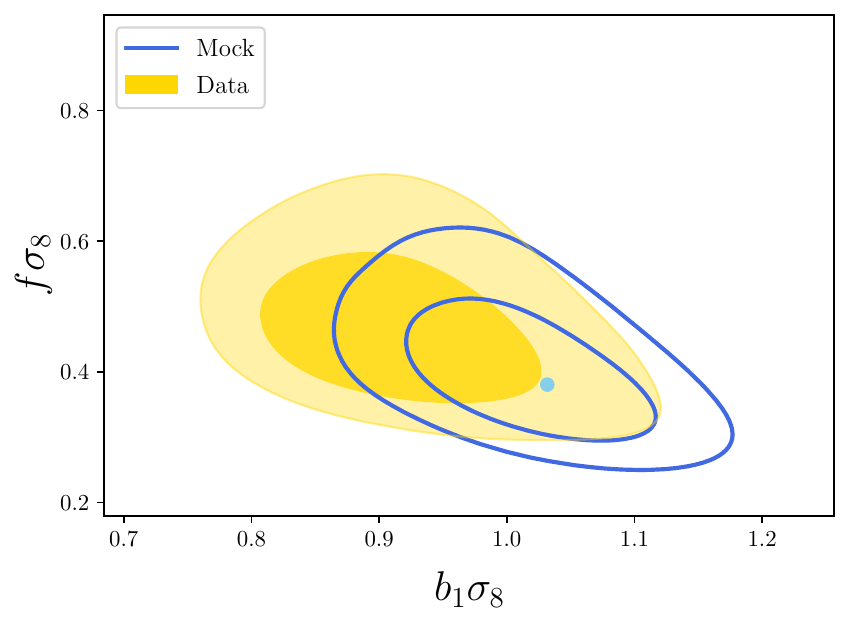}
        \includegraphics[width=0.45\textwidth]{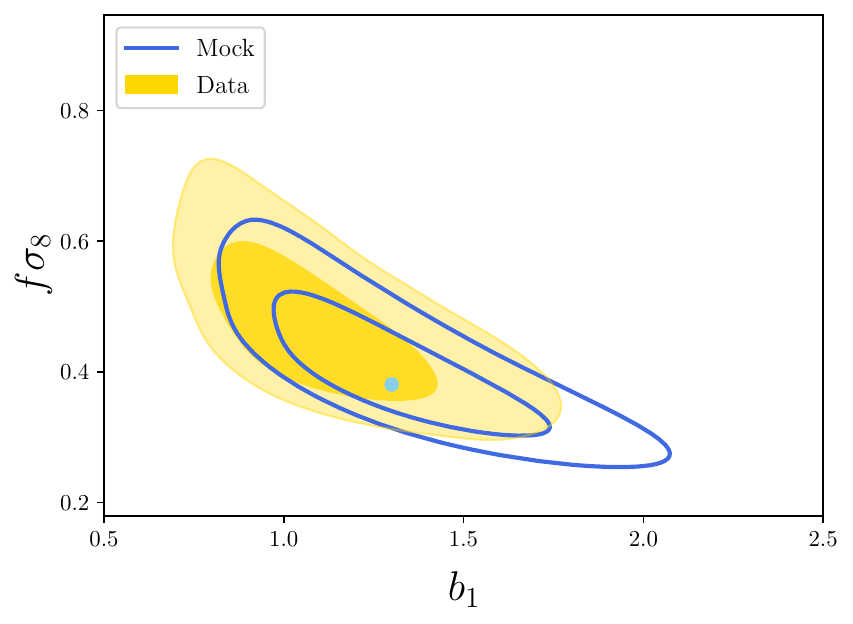}\\
    \includegraphics[width=0.45\textwidth]{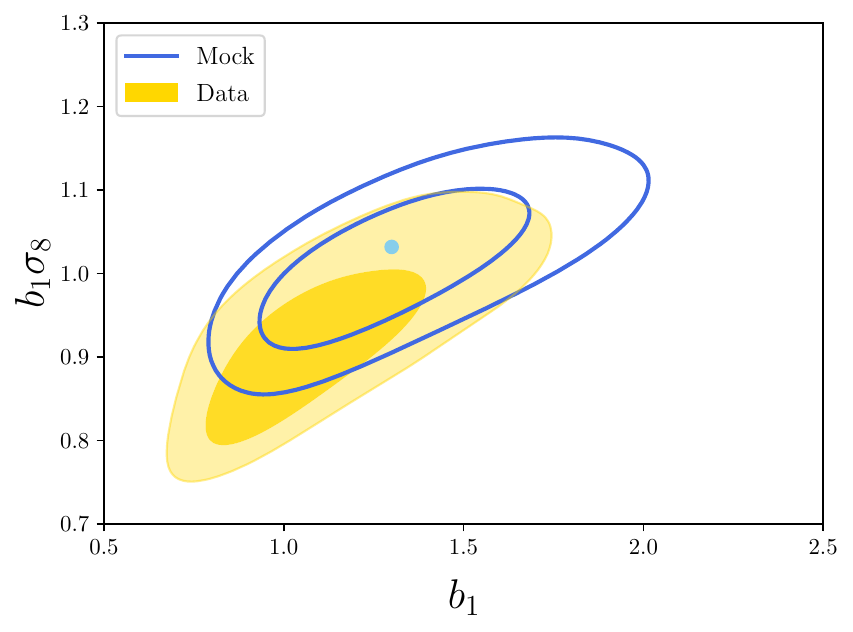}
    \includegraphics[width=0.45\textwidth]{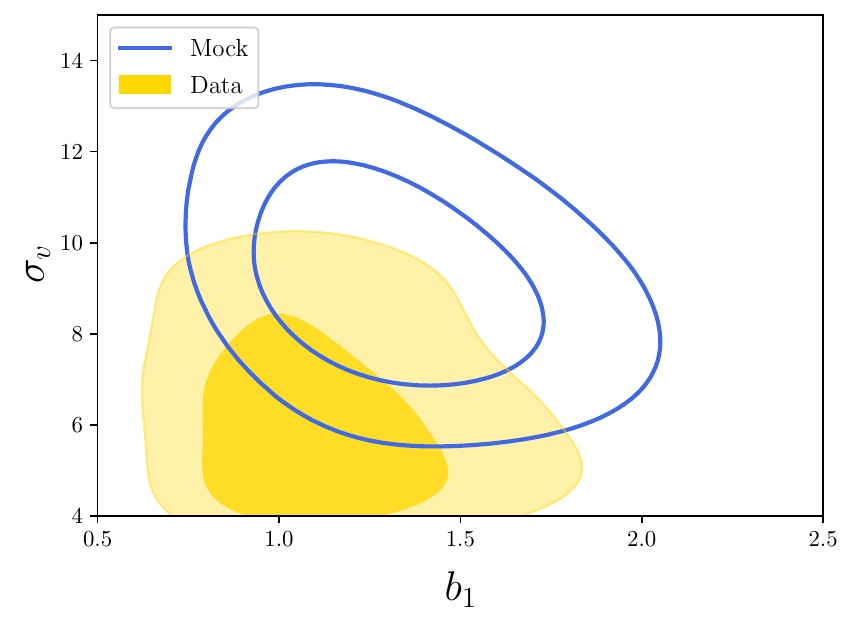}\\
    \caption{Two-dimensional $68/95\%$ marginalised contours on the parameters $(f\sigma_{8}, b_{1}, b_{1}\sigma_{8}, \sigma_{v})$. The empty blue contours are obtained from the mean of the $N=512$ mock samples, and the solid gold are from the SDSS FP data. The pale blue points are the parameter input values used in the HR4 simulation. }
    \label{fig:cont_mock}
\end{figure*}

The most relevant two-dimensional marginalised contours can be found in Figure \ref{fig:cont_mock}, where we present the parameter pairs $(f\sigma_{8},b_{1}\sigma_{8})$, $(b_{1},b_{1}\sigma_{8})$, $(b_{1}, f\sigma_{8})$, $(b_{1},\sigma_{v})$. The colour scheme follows Figure \ref{fig:data} -- blue empty contours represent the posteriors from mean of the mocks and the solid gold contours are our actual results; the cosmological parameters inferred from the SDSS FP data. The pale blue points are the parameter values used to generate the simulation -- $\Omega_{m} = 0.26$, $f = \Omega_{m}^{\gamma} \simeq 0.48$ and $\sigma_{8} = 1/1.26$, $b_{1} = 1.3$, which the blue contours successfully reproduce.

The data favours a lower value of $b_{1}\sigma_{8}$, $b_{1}$ and $\sigma_{v}$ compared to the mocks, but a larger value of $f\sigma_{8}$. The three-way degeneracy between $b_{1}$, $b_{1}\sigma_{8}$ and $f\sigma_{8}$ is apparent from the figures, and is consistent between the mocks and the data. The correlation matrices -- the normalised covariance matrix between the parameters $(b_{1}\sigma_{8}, f\sigma_{8}, b_{1})$ -- are 

\begin{equation} C_{\rm mock} =  \left( \begin{tabular}{ccc}
       1  & -0.51 & 0.68  \\
      & 1 & -0.78 \\
      &  & 1
    \end{tabular} \right) 
\end{equation}
\begin{equation}
    C_{\rm data} =  \left( \begin{tabular}{ccc}
       1  & -0.30 & 0.76  \\
      & 1 & -0.64 \\
      &  & 1
    \end{tabular} \right)
\end{equation}

\begin{figure}
    \centering
    \includegraphics[width=0.45\textwidth]{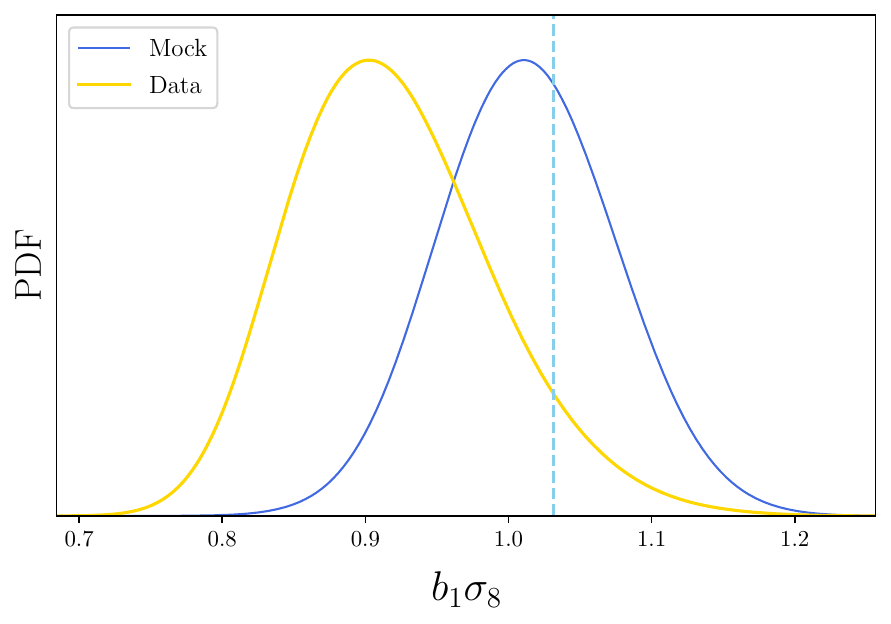}
    \includegraphics[width=0.45\textwidth]{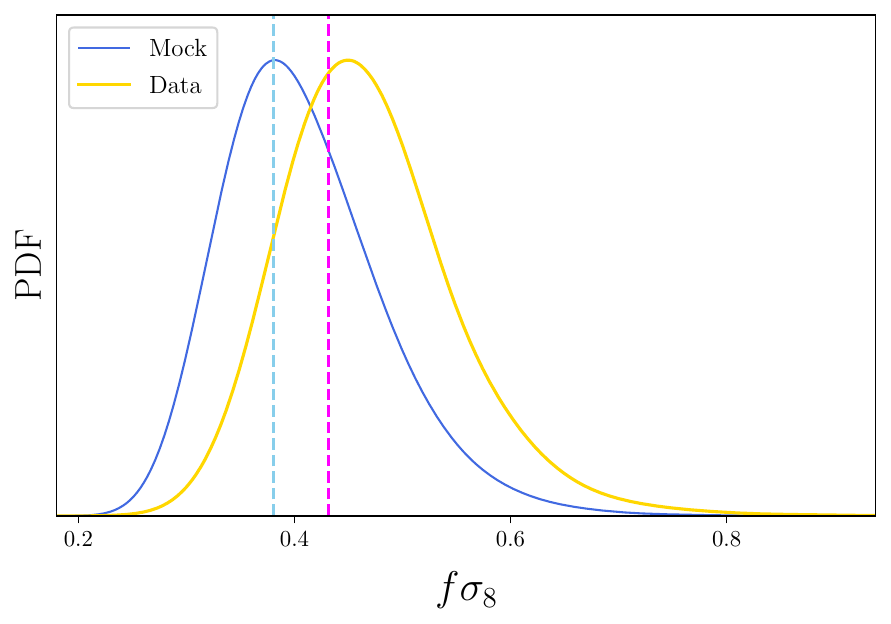}  
    \caption{ One-dimensional marginalised probability distributions (not normalised) for the parameters $b_{1}\sigma_{8}$ (left panel) and $f\sigma_{8}$ (right panel). The color scheme is the same as in Figure \ref{fig:cont_mock}. The pale blue dashed lines are the parameters used in the Horizon Run 4 simulation, and the pink dashed line is the Planck $\Lambda$CDM best fit $f\sigma_{8} = 0.43$, using $\Omega_{m} = 0.3$ and $\sigma_{8} = 0.811$ \citep{Planck:2018vyg}. }
    \label{fig:pdf_mock}
\end{figure}

The one-dimensional posteriors for the parameters $b_{1}\sigma_{8}$ and $f\sigma_{8}$ are presented in Figure \ref{fig:pdf_mock}. The correct input cosmology is recovered from the mean of the mocks (blue vertical dashed lines). The pink dashed line in the right panel is the Planck $\Lambda$CDM best fit $f\sigma_{8} = 0.43$. We note that the peak of the probability distribution function (PDF) of $f\sigma_{8}$ is not exactly the expectation value, since the PDF is skewed towards larger parameter values. This is due to the condition that $f \simeq \Omega_{m}^{6/11} \geq \Omega_{b}^{6/11} \simeq 0.19$, since we implicitly assume that $\Omega_{m}$ cannot be zero due to the presence of baryons -- we fix $\Omega_{b} = 0.048$. Low values of $f\sigma_{8}$ could be realised by allowing $\sigma_{8}$ to be arbitrarily small, and the amplitude of the galaxy power spectrum could in turn be compensated by allowing the linear bias $b_{1}$ to become arbitrarily large, as the matrix indicates, but this is not favoured by the data. The upshot is that $f\sigma_{8}$, which is a derived parameter that is non-linearly related to $\Omega_{m}$, will not generically admit a symmetric posterior.

In Table \ref{tab:2}, we present the expectation values and $68/95\%$ limits on the parameters varied, for the mean of the mocks and the data. The data prefers a value of $f\sigma_{8}$ and $\Omega_{m}$ that is in agreement with the Planck cosmology. The data also very marginally prefers a lower value of $\sigma_{v}$, but this parameter is not well constrained by the data. The value of $b_{1}\sigma_{8}$ is lower than the mocks, which is likely due to the drop in the galaxy density power spectrum on scales $k \sim 0.1 \, h \, {\rm Mpc}^{-1}$.

\begin{figure}
    \centering
    \includegraphics[width=0.45\textwidth]{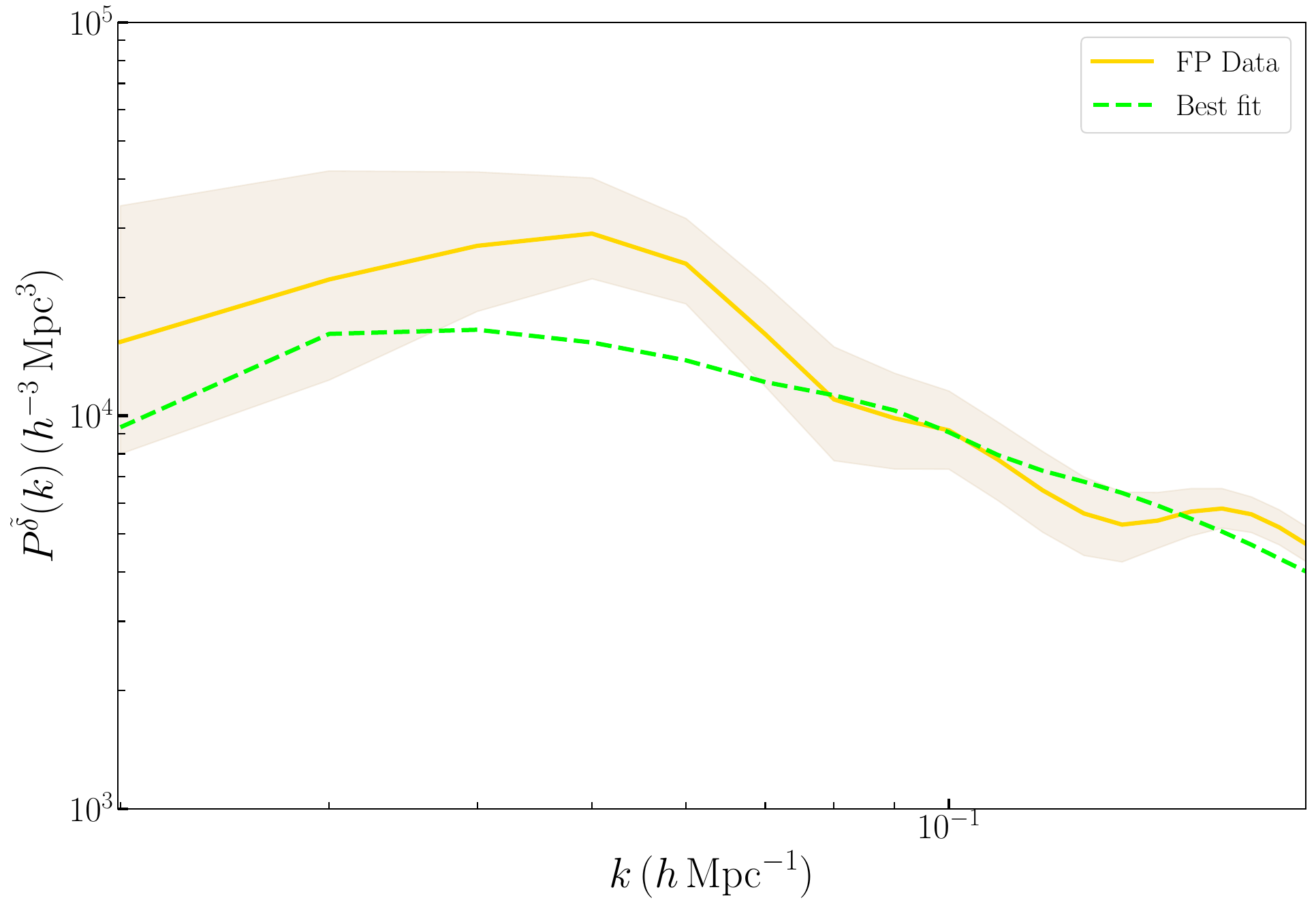}
    \includegraphics[width=0.45\textwidth]{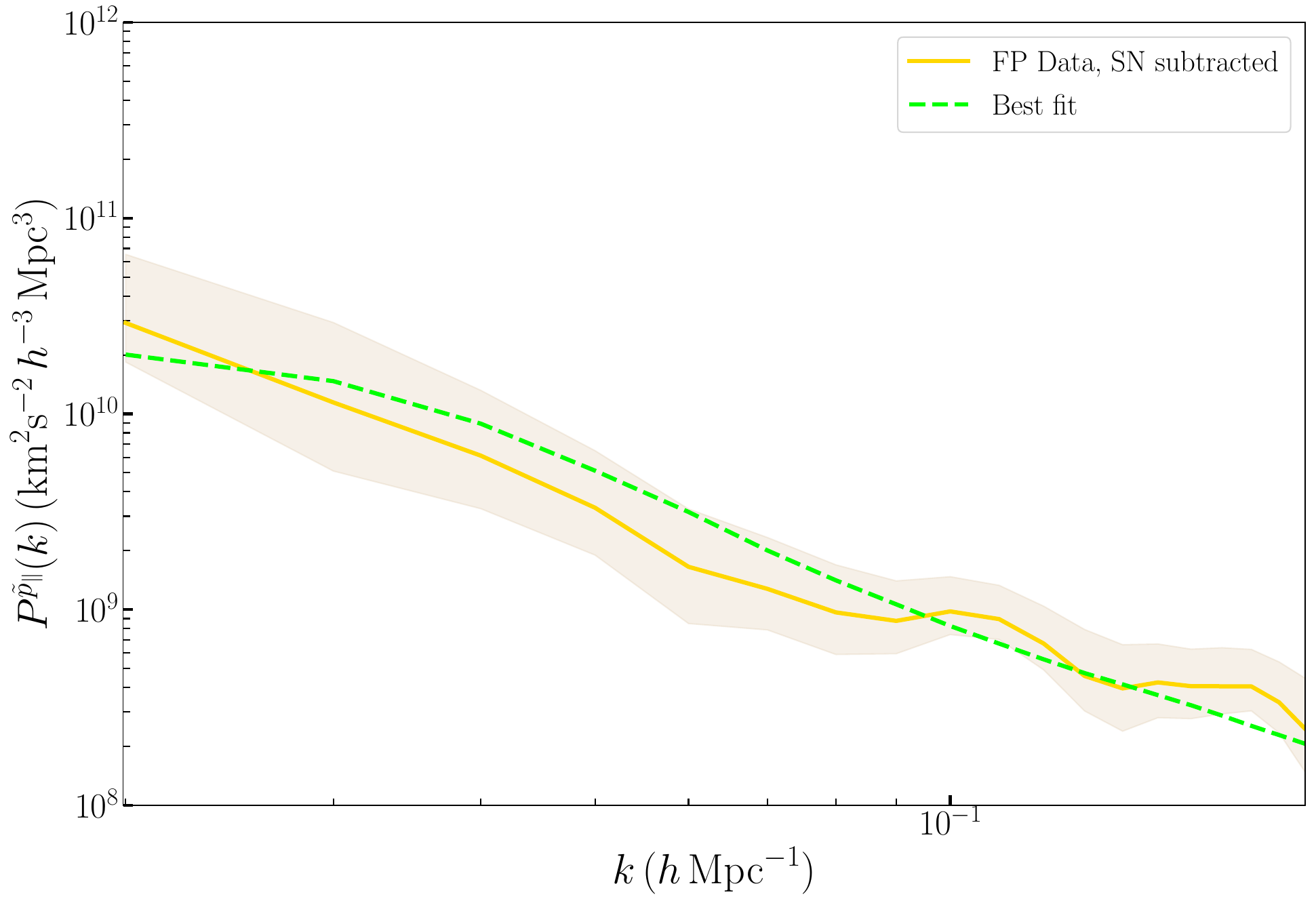}
    \caption{The measured galaxy density (left panel) and momentum (right panel) power spectra (gold lines), and the best fit theoretical power spectra to the SDSS data (green dashed lines). The filled region is the $\pm 1\sigma$ uncertainty from the mock samples. }
    \label{fig:best_fit}
\end{figure}

Finally in Figure \ref{fig:best_fit}, we present the measured galaxy density (left panel) and momentum (right panel) power spectra as solid gold lines, and the best fit theoretical power spectra as green dashed lines. The tan filled region is the $68\%$ confidence region determined from the mocks. The theoretical curve for the galaxy density power spectrum does not fit the large scale modes well (cf. left panel, green curve); the excess of power on large scales relative to the best fit could be mitigated by a lower value of $\Omega_{m}$. We note that an excess of power at scales $k \sim 0.05 \, h \, {\rm Mpc}^{-1}$ was also observed in independent datasets in \citet{Qin:2019axr}. However, we should be careful about performing a chi-by-eye fit to the data since the bins are correlated. The best fit is reasonable; $\chi^{2} = 37.2$ for the number of degrees of freedom $N_{\rm dof} = 34$ ($38$ data points, $4$ free parameters).

\begin{table*}
 \begin{tabular}{||c c c c c c ||}
 \hline  
 Parameter \, &  $f\sigma_{8}$ \, & $b_{1}\sigma_{8}$ \, &  $\Omega_{m}$ \,  & $\sigma_{v}$ \,  & $\chi^{2}$ \, \\ [0.5ex] 
 \hline 
  \, & \,  & \, &  \,  &  \,  &  \, \\ 
 Mock  \, &  $0.407_{-0.073-0.124}^{+0.073+0.177}$ \, & $1.014_{-0.061-0.115}^{+0.061+0.122}$ \, &  $0.306_{-0.062-0.095}^{+0.059+0.142}$ \, & $9.08_{-1.54-2.74}^{+1.64+3.18}$ \, & 4.0 \,   \\
 \, & \,  & \, &  \,  &  \, & \,  \\ 
    Data \, &  $0.471_{-0.080-0.141}^{+0.077+0.192}$ \, & $0.920_{-0.070-0.123}^{+0.070+0.154}$ \, &  $0.341_{-0.067-0.118}^{+0.074+0.173}$ \, &  $6.12_{-1.64-2.04}^{+1.60+3.68}$ \,    & 37.2 \, \\
 \, & \,  & \, &  \,  &  \, & \,  \\ 
 \hline
\end{tabular}
\caption{\label{tab:2} The best fit and $68/95\%$ ranges for the most relevant parameters studied in this work.   }
\end{table*}

\section{Discussion} 
\label{sec:disc}

In this work we extracted the galaxy density and momentum power spectrum from an ETG subset of the SDSS main galaxy sample, using FP information to infer velocities. We compared the measurements to power spectra derived using perturbation theory up to third order, which is formulated using the $\Lambda$CDM model. After testing our analysis on mock galaxy catalogs, we arrive at a constraint of $b_{1}\sigma_{8} = 0.920_{-0.070}^{+0.070}$ and $f\sigma_{8} = 0.471_{-0.080}^{+0.077}$ at a mean redshift $\bar{z} = 0.04$.

Our analysis is consistent with other measurements of the same parameters in the literature \citep{Fisher:1993ye,Peacock:2001gs,Blake:2011rj,Beutler:2012px,delaTorre:2013rpa,BOSS:2016wmc,Gil-Marin:2020bct,Reid:2012sw,Bautista:2020ahg,Qin:2019axr}. The momentum power spectrum is a sum of various two point functions, and hence sensitive to a combination of $b_{1}$, $\sigma_{8}$, $f$, $\Omega_{m}$, not simply to $f\sigma_{8}$. For this reason, we tried to simultaneously fit for $b_{1}$, $b_{1}\sigma_{8}$ and $\Omega_{m}$, relying on the well-developed theoretical framework of the $\Lambda$CDM and treated $f\sigma_{8}$ as a derived parameter. It is common practice in the literature to treat $f\sigma_{8}$ as a free parameter and then compare it to theoretical predictions. However, to extract this quantity from the galaxy density and momentum fields, we restricted our analysis entirely to the confines of the $\Lambda$CDM model. This is because we are extracting information from both the amplitude and shape of the power spectra. We expect that any modified gravity/cosmological model would not only change $f\sigma_{8}$, but also the kernels used in perturbation theory, and introduce additional scales -- such as scalar field masses -- that will further modify the shape of the power spectrum. In this work we did not build the theoretical templates for such a general class of gravity models. Still, any deviation from the $\Lambda$CDM model could be detected by comparing our derived $f\sigma_{8}$ to values inferred from other probes. 

Furthermore, we found that $f\sigma_{8}$ does not yield a symmetric marginalised posterior when fitting the $\Lambda$CDM model to the data, since $f$ is a nonlinear function of $\Omega_{m}$ that is in turn bounded by the condition $\Omega_{m} \geq \Omega_{b}$. In this case, taking $f\sigma_{8}$ as a derived parameter and using the $\Lambda$CDM expectation $f \simeq \Omega_{m}^{6/11}$ is our preferred choice. By allowing $f\sigma_{8}$ to be an independent, free parameter with uniform prior, we might be making a subtly different assumption that could impact our interpretation of the result. The skewness of the $f\sigma_{8}$ posterior will be particularly pronounced at lower best fit values. Although combinations of measurements hint at a growth rate that departs from the $\Lambda$CDM model \citep{Nguyen:2023fip}, it may be difficult to exactly ascertain the significance of any anomalous measurements of this parameter due to the non-Gaussian nature of the posterior. 

We also comment on the non-linear velocity dispersion and its effect on the power spectrum. The Fingers of God produce a very pronounced effect on the momentum power spectrum, on surprisingly large scales. This is discussed further in the appendix, but we find an order $\sim {\cal O}(20-30\%)$ effect on scales $k < 0.1 h \, {\rm Mpc}^{-1}$. Non-linear stochastic velocities generate a large decrease in power on large scales, followed by an increase in power on smaller scales. Such behaviour has already been noted in the literature \citep{Koda:2013eya, Dam:2021fff}. This phenomena will depend on the galaxy sample under consideration -- central galaxies will be less affected by non-linear velocities. Understanding the Finger of God effect on the velocity and momentum statistics remains an interesting topic of future study. In this work, we used a single free parameter to model this effect. Given the quality and volume of data, this was sufficient. Future analysis will require more consideration. 

The low redshift SDSS FP velocity data is consistent with other cosmological probes such as the CMB. However, the volume of data is modest and statistical uncertainties are large; the current data does not have strong model discriminatory power. In addition, we find that non-linear velocity dispersion strongly affects the momentum power spectrum, which forces us to introduce phenomenological prescriptions to model the effect. This philosophy is counter to the standard cosmology orthodoxy, which is to perform perturbation theory on an FLRW background. Certainly, there is room for alternative spacetime metrics to fit the data with equal precision, and this idea will be pursued in the future. The SDSS FP data is at or below the homogeneity scale assumed within the standard $\Lambda$CDM model, and searching for alternative prescriptions of the low redshift Universe is an interesting direction of future study.

\acknowledgements
SA and MT are supported by an appointment to the JRG Program at the APCTP through the Science and Technology Promotion Fund and Lottery Fund of the Korean Government, and were also supported by the Korean Local Governments in Gyeongsangbuk-do Province and Pohang City. SEH was supported by the project \begin{CJK*}{UTF8}{mj}우주거대구조를 이용한 암흑우주 연구\end{CJK*} (``Understanding Dark Universe Using Large Scale Structure of the Universe''), funded by the Ministry of Science. JK was supported by a KIAS Individual Grant (KG039603) via the Center for Advanced Computation at Korea Institute for Advanced Study. We thank the Korea Institute for Advanced Study for providing computing resources (KIAS Center for Advanced Computation Linux Cluster System). 

Funding for the SDSS and SDSS-II has been provided by the Alfred P. Sloan Foundation, the Participating Institutions, the National Science Foundation, the U.S. Department of Energy, the National Aeronautics and Space Administration, the Japanese Monbukagakusho, the Max Planck Society, and the Higher Education Funding Council for England. The SDSS Web Site is http://www.sdss.org/. The SDSS is managed by the Astrophysical Research Consortium for the Participating Institutions. The Participating Institutions are the American Museum of Natural History, Astrophysical Institute Potsdam, University of Basel, University of Cambridge, Case Western Reserve University, University of Chicago, Drexel University, Fermilab, the Institute for Advanced Study, the Japan Participation Group, Johns Hopkins University, the Joint Institute for Nuclear Astrophysics, the Kavli Institute for Particle Astrophysics and Cosmology, the Korean Scientist Group, the Chinese Academy of Sciences (LAMOST), Los Alamos National Laboratory, the Max-Planck-Institute for Astronomy (MPIA), the Max-Planck-Institute for Astrophysics (MPA), New Mexico State University, Ohio State University, University of Pittsburgh, University of Portsmouth, Princeton University, the United States Naval Observatory, and the University of Washington.

Some of the results in this paper have been derived using the healpy and HEALPix package.

\bibliographystyle{aasjournal}
\bibliography{biblio}

\appendix
\section{Bias Parameters}

We reconstruct the bias parameters of the Horizon Run 4 mock catalog. We fix the cosmological parameters to their input values $\Omega_{m} = 0.26$, $\Omega_{b} = 0.048$, $h = 0.72$, $n_{\rm s} = 0.96$, $\sigma_{8} = 1/1.26$, then fit the theoretical model contained in Section \ref{sec:theory} to infer $b_{1}$, $b_{2}$, $b_{3, nl}$, $b_{s}$. We perform this test using the entire Horizon Run 4 box, measuring the real space galaxy density power spectrum thus eliminating all redshift space contributions to the theoretical power spectrum. We estimate the statistical uncertainty on the measurement by assuming a Gaussian covariance -- 

\begin{equation} C_{ij} = {(2\pi)^{3} \over V} {[P^{\delta}(k_{i})]^{2} \over 2\pi k_{i}^{2} \Delta  k} \delta_{ij}
\end{equation} 

\noindent where $\delta_{ij}$ is the Kronecker delta, indicating a diagonal covariance between $i$ and $j$ Fourier bins. We use the same $N_{k} = 19$ Fourier bins as in the main body of the paper, so $\Delta k = 0.01$ and $V$ is the volume of the data. Rather than taking $V = (3150 \, h^{-1} \, {\rm Mpc})^{3}$, we scale this parameter such that the $\chi^{2}$ per degree of freedom is approximately unity, when we fit the bias model to the data. This requires $V \simeq (1200 \, h^{-1} \, {\rm Mpc})^{3}$. We have checked that this procedure does not change the best fit values that we obtain from the entire snapshot box, only the width of the posteriors of the bias parameters. We do not expect the width of the posteriors to be meaningful since we are using a very approximate Gaussian, diagonal covariance matrix, but we are only interested in the best fit values, which are robust. We minimize the $\chi^{2}$ function 

\begin{equation} \chi^{2} =  (P^{\delta}_{i} -  \grave{P}^{\delta}_{i})C^{-1}_{ij}(P^{\delta}_{j} - \grave{P}^{\delta}_{j}) , \end{equation} 

\noindent where $P^{\delta}_{i}$ is the measured value of the real space galaxy power spectrum in the $i^{\rm th}$ Fourier bin, and $\grave{P}^{\delta}_{i}$ is the non-linear theoretical power spectrum in real space -- that is, the power spectrum (\ref{eq:th1}) after fixing all redshift space contributions to zero.

\begin{figure}
    \centering
    \includegraphics[width=0.45\textwidth]{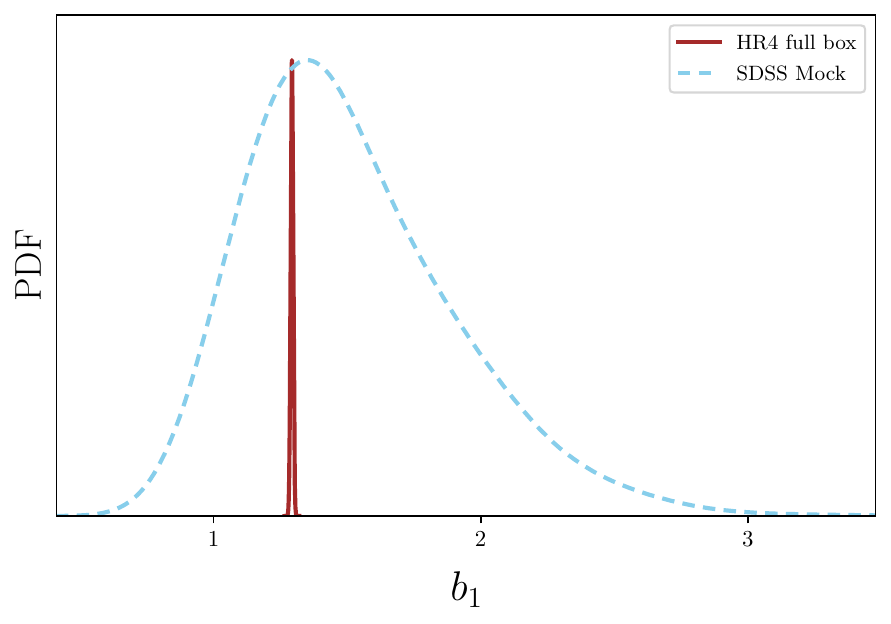}
    \includegraphics[width=0.45\textwidth]{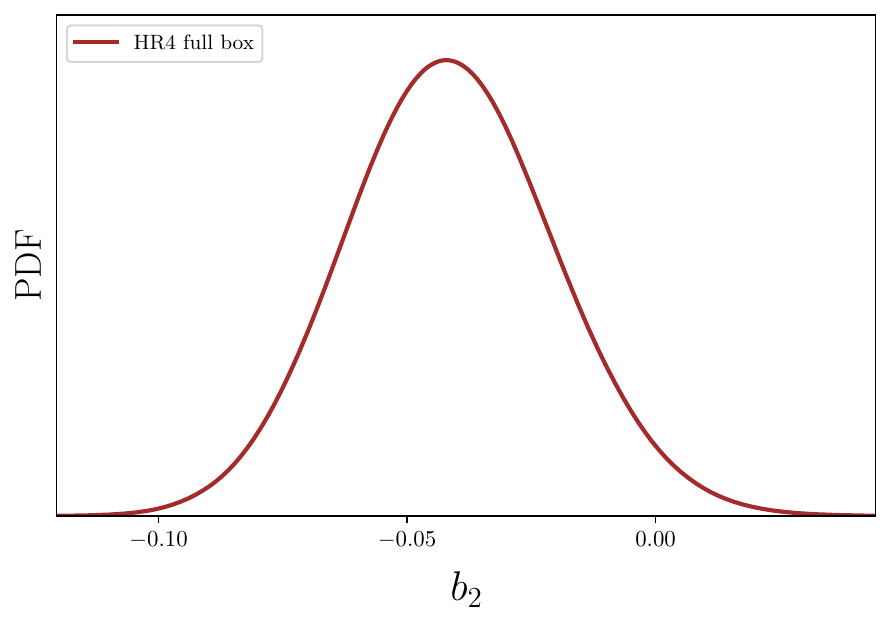}
    \caption{Marginalised 1D posterior distributions of $b_{1}$ (left panel) and $b_{2}$ (right panel) from the full Horizon Run 4 snapshot box (brown), and the mock SDSS data (blue dashed). }
    \label{fig:app_IIb_bias}
\end{figure}

In Figure \ref{fig:app_IIb_bias} we present the one-dimensional marginalised PDF's of the parameters $b_{1}$ (left panel) and $b_{2}$ (right panel). In the left panel, the brown curve is the posterior obtained from the full, real-space snapshot box in this Appendix. The blue dashed line is the PDF of $b_{1}$ inferred from the mock data in the main body of the text. It is clear that the mock SDSS data cannot constrain this parameter, as the volume is too small and statistical uncertainties too large. Still, there is consistency between our mocks and the `true' value inferred from the full box, $b_{1} = 1.3$. Similarly, in the right panel we present $b_{2}$ constraints from the full box. When generating the parameter constraints in the main body of the text, we have tried fixing $b_{2} = -0.05$ using the HR4 `true' value and also allowed this parameter to vary over $-1 \leq b_{2} \leq 1$, finding that variation of $b_{2}$ does not change the posteriors on $f\sigma_{8}$ and $b_{1}\sigma_{8}$ when fitting the cosmological model to the data. Finally, we find that even when using the entire snapshot box, we cannot infer a constraint on $b_{3, nl}$ or $b_{s}$ over the prior range $-10 < b_{3, nl}, b_{s} < 10$, using the Fourier modes $0.02 \, h \, {\rm Mpc}^{-1} < k < 0.2 \, h \, {\rm Mpc}^{-1}$, finding constraints of $b_{3, nl} = -0.54 \pm 4.5$ and $b_{s} = 0.23 \pm 5.9$. We fix these parameters in the main body of the text. 

%\subsubsection*{{\rm B -- Redshift Space without the Finger of God Effect}} 
\section{Redshift Space without the Finger of God Effect}

The Horizon Run 4 mock catalogs contain information on both galaxy position and velocity, and the host halo velocities. With this information we can construct different redshift space fields from the data, and disentangle cosmological redshift space distortion from non-linear phenomenon such as the Finger of God effect. Specifically, if we take the galaxy catalogs and assign to them their host halo velocities, we derive a momentum field in which the peculiar velocity between host halo and galaxy is removed, mitigating the Finger of God effect.

In this section we measure three distinct momentum and galaxy density power spectra from the Horizon Run 4 snapshot box. They are distinguished by the observable that we measure -- density field and momentum field -- and the velocity that we use to generate redshift space distortion. The six mock observables are summarized in Table \ref{tab:app_1}. The $g, h$ subscripts on the power spectra $P_{g,h}$ denote that galaxy/halo velocities are used to correct the galaxy positions when generating redshift space fields. The galaxy velocity is always used as the observable when constructing the momentum field $p_{\parallel}$ and $\tilde{p}_{\parallel}$. 

We use the full snapshot box in this section, and take ${\bf e}_{\parallel} = {\bf e}_{z}$. The quantities $P^{\tilde{\delta}}_{g}$ and $P^{\tilde{p}_{\parallel}}_{g}$ are directly measurable from actual data, although by using the galaxy velocities one could also reconstruct the real space fields $P^{\delta}$ and $P^{p_{\parallel}}$ \citep{Park:2000rc,Park:2005zb}, albeit with large scatter due to the FP uncertainty. The halo redshift space quantities $P^{\tilde{\delta}}_{h}$ and $P^{\tilde{p}_{\parallel}}_{h}$ are mock constructs that allow us to disentangle non-linear stochastic peculiar velocities between galaxies occupying the same host halo and the larger scale velocities that contain cosmological information.

\begin{table}
\begin{center}
 \begin{tabular}{||c c c c ||}
 \hline 
 RS velocity correction $\rightarrow$ \, & Real Space \, & Halo Velocity \, &  Galaxy Velocity \,   \\
 Observable $\downarrow$ \, & \, & \, & \,  \\  
  \hline 
 $\delta$ \, & $P^{\delta}$ \, & $P^{\tilde{\delta}}_{\rm h}$ \, & $P^{\tilde{\delta}}_{\rm g}$ \,   \\   
   $p_{\parallel}$ \, & $P^{p_{\parallel}}$ \, & $P^{\tilde{p}_{\parallel}}_{\rm h}$ \, & $P^{\tilde{p}_{\parallel}}_{\rm g}$ \,  \\ 
  \hline
\end{tabular}
\caption{\label{tab:app_1} The list of the power spectra considered. Different columns mean different spaces (real space, redshift space sourced by host halo velocities, and redshift space sourced by galaxy velocities). Different rows mean different quantities from which the power spectrum is calculated.
}
\end{center} 
\end{table}

\begin{figure}
    \centering
    \includegraphics[width=0.9\textwidth]{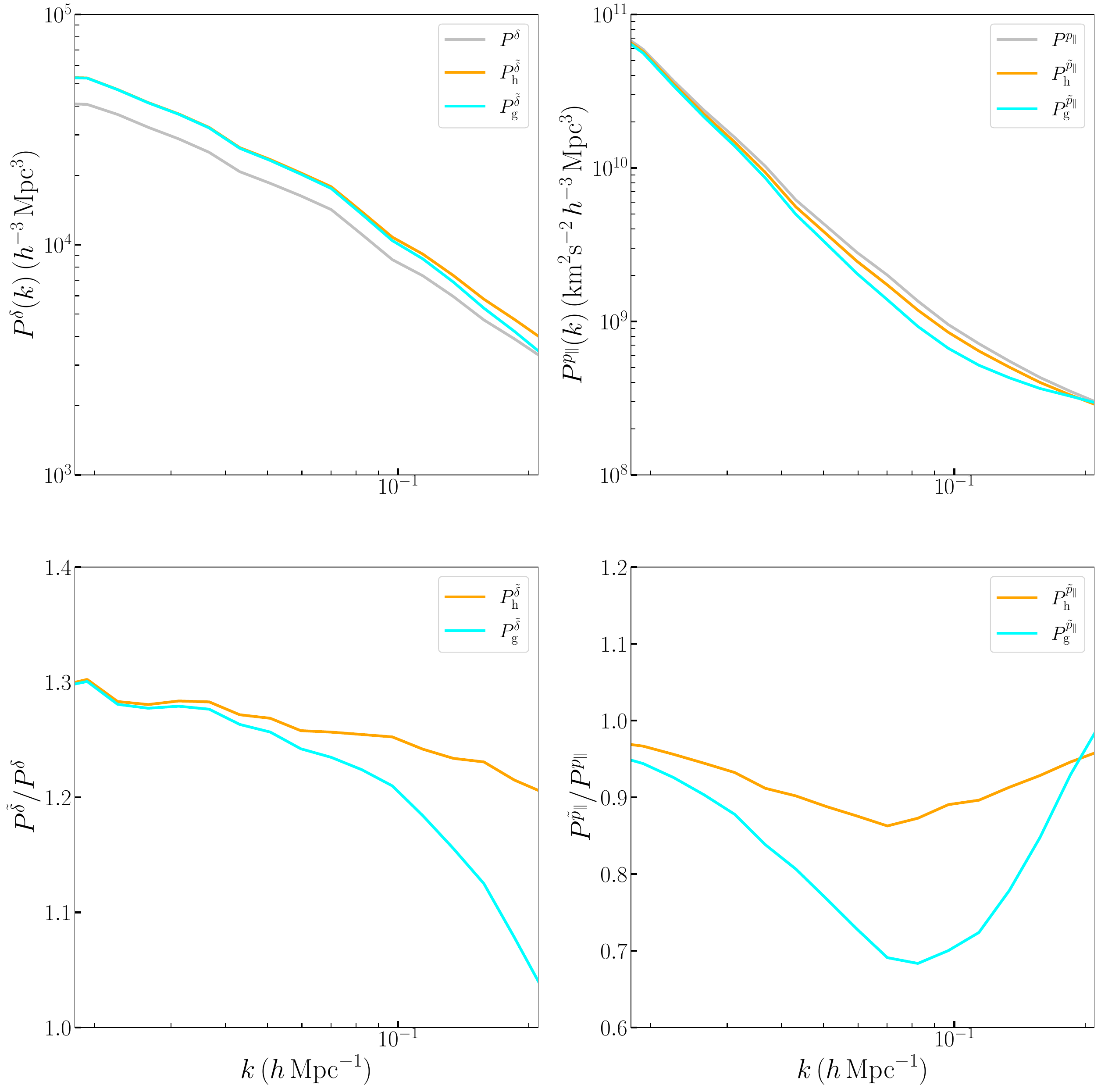}
    \caption{Galaxy density (top left) and momentum (top right) power spectra extracted from the Horizon Run 4, $z=0$ snapshot box. The grey/orange/cyan lines are the statistics in real space, halo-source redshift space and galaxy-sourced redshift space. In the lower panels we present the ratio of the two redshift space power spectra and the real space counterpart.  }
    \label{fig:app_dd_pp}
\end{figure}

\noindent In Figure \ref{fig:app_dd_pp} we present the galaxy density (top left) and momentum (top right) power spectra measured from these multiple fields. In the both panels, the grey, orange, cyan lines are the  power spectra in real, halo-sourced and galaxy-sourced redshift space. The lower left/right panels contain the ratios $P^{\tilde{\delta}}_{\rm h}/P^{\delta}$, $P^{\tilde{\delta}}_{\rm g}/P^{\delta}$ and  $P^{\tilde{p}_{\parallel}}_{\rm h}/P^{p_{\parallel}}$, $P^{\tilde{p}_{\parallel}}_{\rm g}/P^{p_{\parallel}}$  (cyan/orange lines) respectively.

The galaxy density power spectrum behaves as expected. The effect of redshift space distortion is to increase the amplitude of the power spectrum on all scales, and there is an additional suppression of power on small scales due to the stochastic velocities of bound structures (cf. cyan curve, bottom panel). The orange curve, which is a hypothetical redshift space using the host halo velocity to correct galaxy positions, almost entirely removes the small scale suppression due to the Fingers of God. 

The effect of redshift space distortion on the momentum power spectrum is different. On large scales, all three power spectra approach a common value -- although there is a $\sim 5\%$ discrepancy between real and redshift space momentum power spectra even on scales $k \sim 0.01 h \, {\rm Mpc}^{-1}$. The finger of God effect has a particularly large impact on $P^{\tilde{p}_{\parallel}}$ (cf. cyan line, right panels). It generates a suppression of power as large as $\sim 30\%$ on scales $k \sim 0.05 h \, {\rm Mpc}^{-1}$, with a subsequent increase in power on small scales. This behaviour has been observed in previous works \citep{Koda:2013eya}, and is typically modelled using a series of phenomenological prescriptions such as exponential damping and a shot noise contribution. However, even when using the halo velocities to generate redshift space fields there is a suppression in power (cf. orange lines, right panels), which suggests that both nonlinear perturbative effects and the Finger of God contribution act to suppress the momentum power spectrum. This was also noted in \citet{Dam:2021fff}, who studied the velocity two-point statistics.

\section{Ancillary Information}

\begin{figure*}
 \begin{center}
  \includegraphics[width=0.5\textwidth]{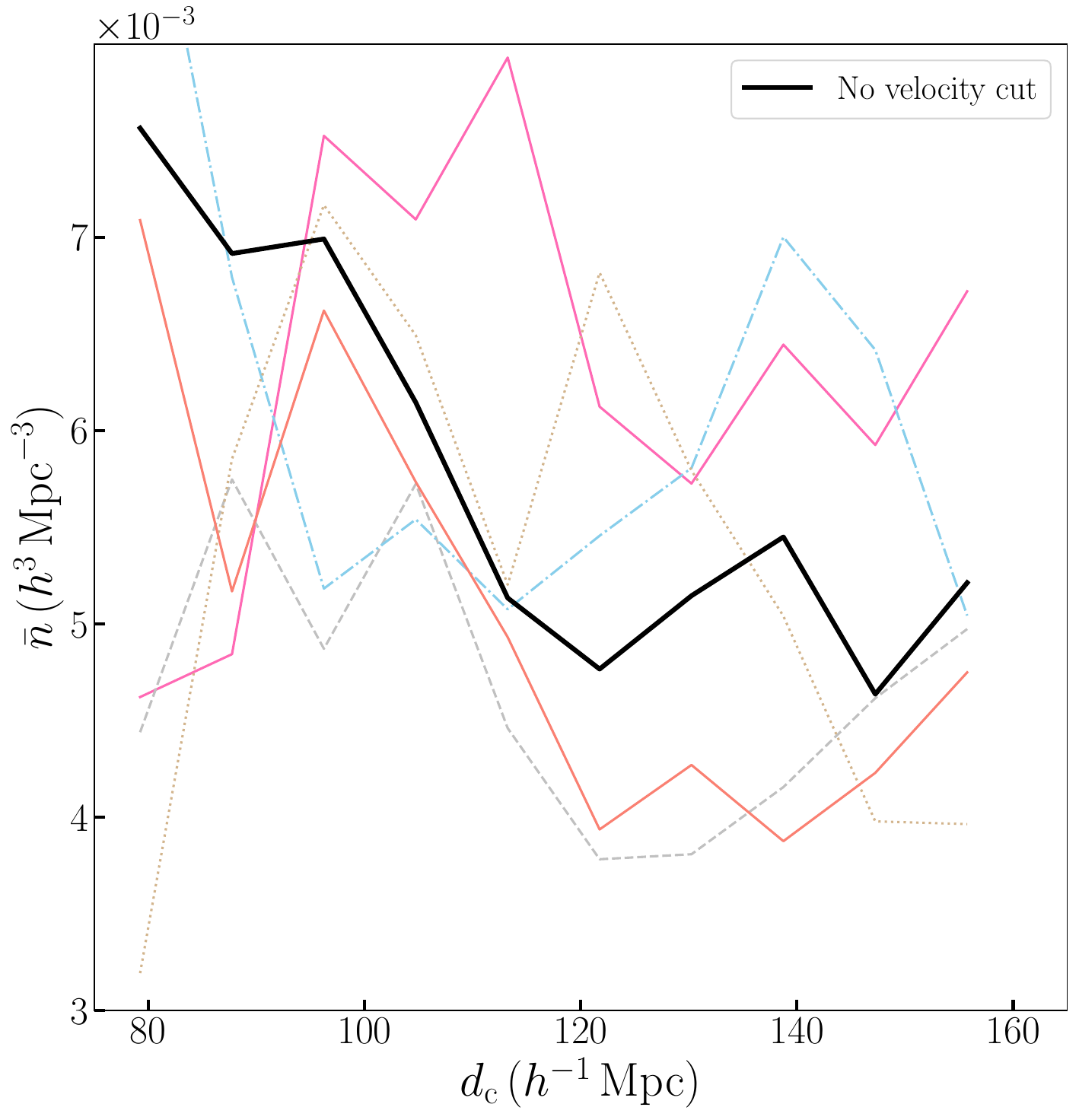}
 \end{center}
 \caption{\label{fig:app_nbar} Some random $\bar{n}$ vs $z$ curves obtained from mock galaxy samples, with the SDSS data (solid black curve). We observe that the variation of $\bar{n}$ within the SDSS data is typical.
}
\end{figure*}

In this appendix, we present some ancilliary information regarding the mocks. In Figure \ref{fig:app_nbar} we present some $\bar{n}$ vs $z$ curves from random mock samples, overlaid with the SDSS data. We see that the variation of the density within the SDSS data is consistent with typical variations observed within the mocks.

\begin{figure*}
 \begin{center}
  \includegraphics[width=0.95\textwidth]{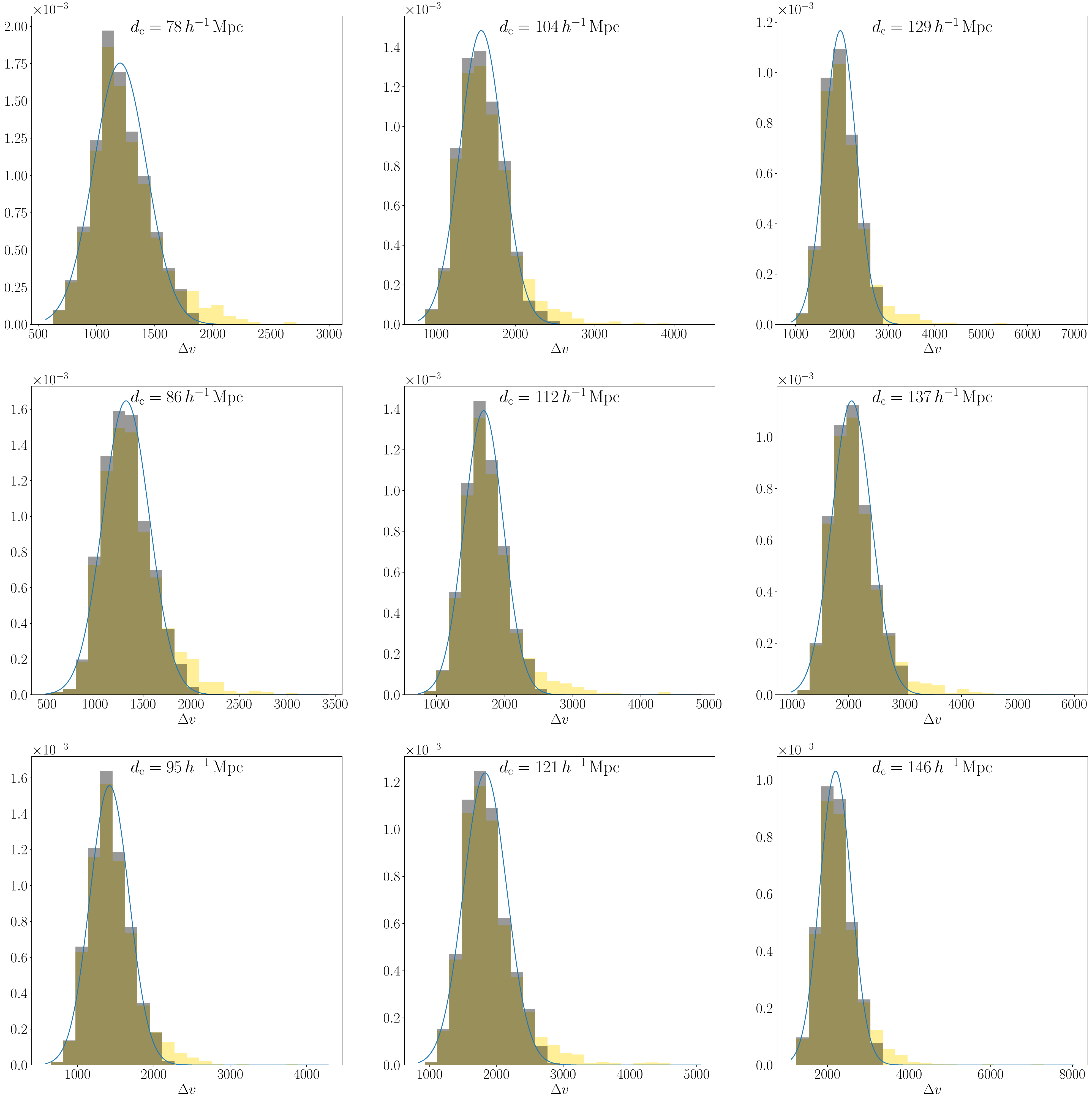}
 \end{center}
 \caption{\label{fig:histograms} After binning the velocity uncertainties into nine redshift bins, we fit a Gaussian distribution to the cut data (blue curves). The dark bars are the cut SDSS data and yellow bars the full sample. The choice of cut approximately Gaussianizes the velocity uncertainty in each redshift bin.
}
\end{figure*}

In Figure \ref{fig:histograms} we decompose the SDSS FP galaxies into nine redshift bins and generate histograms of the velocity uncertainty in each bin. The gold bars represent the entire sample, and the black bars the remaining galaxies after making a $5\%$ $\Delta z$ cut as described in the main body of the text. The blue curves are Gaussian distributions fitted to each black histogram. We observe that the $\Delta v$ cut largely Gaussianizes the velocity uncertainty, which justifies our treatment of the FP noise as a Gaussian white noise component. In Table \ref{tab:app_2} we present the sample mean, standard deviation, skewness and kurtosis of the data in the nine redshift bins, before (after) the cut is made. The cut significantly reduces the non-Gaussian skewness and kurtosis of the samples.

\begin{table*}
 \begin{tabular}{||c c c c c ||}
 \hline  
 Bin $d_{\rm cm} \, (h^{-1} \, {\rm Mpc})$ \, &  Mean (${\rm km \, s^{-1}}$) \, & Std. Deviation (${\rm km \, s^{-1}}$) \, &  Skewness \,  & Kurtosis \,  \\ [0.5ex] 
 \hline 
 $78$  \, &  $1248 \, (1203)$ \, & $292 \, (227)$ \, &  $1.09 \, (0.31)$ \, & $1.96 \, (-0.20)$ \,    \\
 $86$  \, &  $1373 \, (1322)$ \, & $313 \, (242)$ \, &  $1.13 \, (0.27)$ \, & $2.52 \, (-0.16)$ \,    \\
 $95$  \, &  $1459 \, (1418)$ \, & $322 \, (256)$ \, &  $1.30 \, (0.29)$ \, & $4.24 \, (-0.23)$ \,    \\
 $104$  \, &  $1625 \, (1571)$ \, & $353 \, (269)$ \, &  $1.39 \, (0.30)$ \, & $4.27 \, (-0.08)$ \,    \\
 $112$  \, &  $1760 \, (1691)$ \, & $406 \, (286)$ \, &  $1.70 \, (0.24)$ \, & $5.76 \, (+0.06)$ \,    \\
 $121$  \, &  $1897 \, (1831)$ \, & $437 \, (322)$ \, &  $1.64 \, (0.36)$ \, & $5.23 \, (-0.16)$ \,    \\
 $129$  \, &  $2042 \, (1964)$ \, & $482 \, (341)$ \, &  $1.92 \, (0.29)$ \, & $8.53 \, (-0.26)$ \,    \\
 $137$  \, &  $2119 \, (2054)$ \, & $469 \, (349)$ \, &  $1.65 \, (0.38)$ \, & $5.24 \, (-0.24)$ \,    \\
 $146$  \, &  $2278 \, (2199)$ \, & $523 \, (387)$ \, &  $1.82 \, (0.31)$ \, & $8.63 \, (-0.30)$ \,    \\
 \hline
\end{tabular}
\caption{\label{tab:app_2} The sample mean, standard deviation, skewness and kurtosis of $\Delta v$ in nine redshift bins, before and after the $\Delta v$ cut. The numbers in the brackets indicate the values after the cut has been applied. The sample is Gaussianized by our choice.    }
\end{table*}

\begin{figure*}
 \begin{center}
  \includegraphics[width=0.95\textwidth]{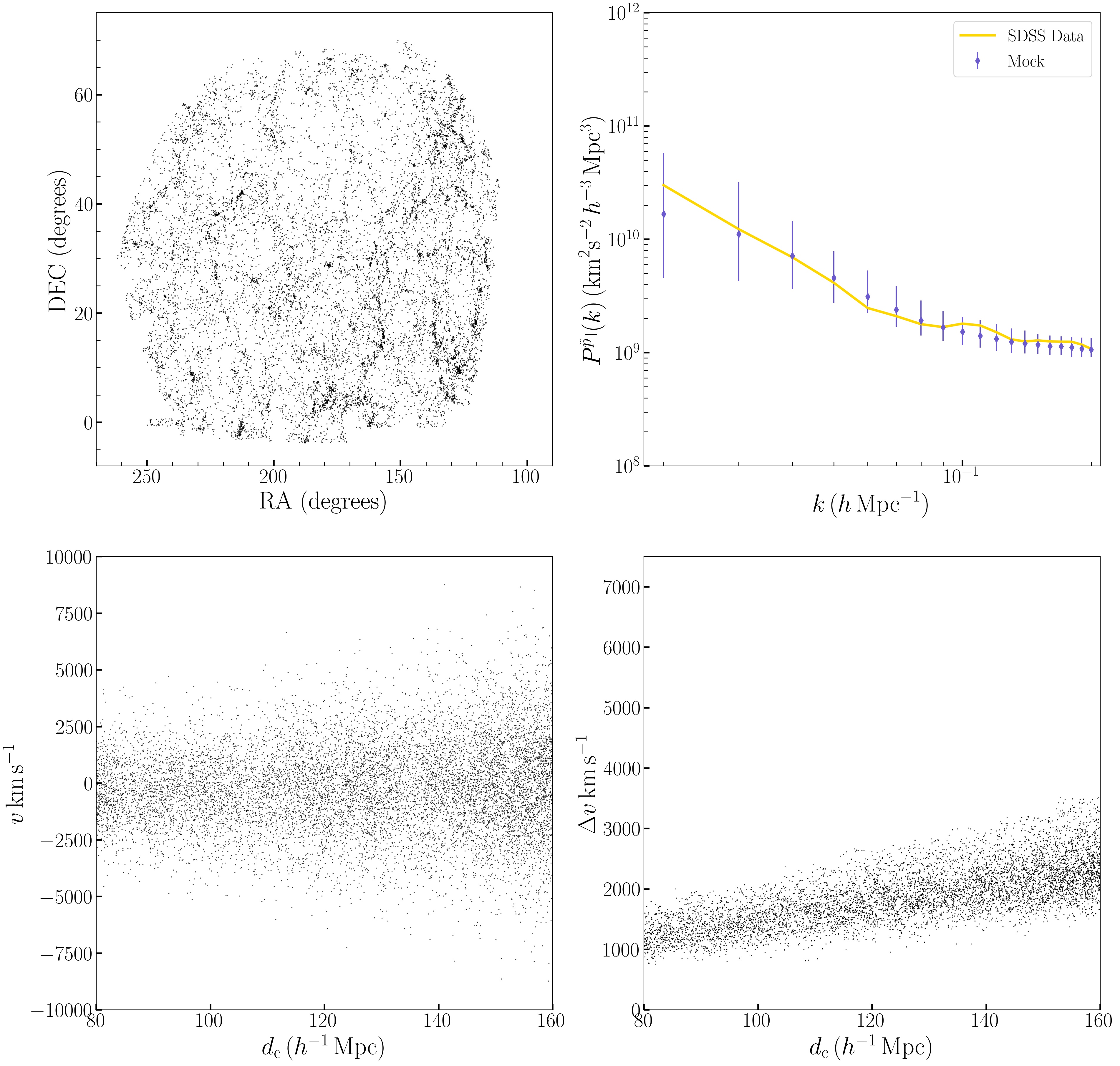}
 \end{center}
 \caption{\label{fig:mocks} Angular distribution of a typical mock catalog used in this work (upper left), the velocity as a function of comoving distance after a velocity uncertainty has been added to each galaxy (lower left) and the velocity uncertainty as a function of comoving distance (lower right panel). In the top right panel we present the median and $68\%$ range of the momentum power spectra extracted from the $512$ mock catalogs, and the solid yellow line is the power spectrum extracted from the SDSS. 
}
\end{figure*}

Finally, we check that the mock data can approximately mimic the SDSS catalog at the level of the field and subsequent power spectrum extraction, by introducing a velocity uncertainty to the mock galaxies. For each of the $512$ mock realisations, to each galaxy we add a velocity uncertainty $\Delta v$. This is drawn from a Gaussian, of zero mean and width equal to the velocity uncertainty of the SDSS galaxy closest to the mock in comoving distance\footnote{We use the cut SDSS catalog for this purpose, with the $5\%$ of highest $\Delta v$ galaxies removed.}. In Figure \ref{fig:mocks} we present the (RA, dec) of a typical mock (upper left panel), the velocity against comoving distance (lower left panel) and the velocity error against comoving distance (lower right panel). In the upper right panel we present the median and $68\%$ limits of the momentum power spectra extracted from these mocks, along with the SDSS measurement. We see that the mocks qualitatively reproduce the power spectrum of the data. Also, the $v$--$z$ and $\Delta v$--$z$ plots (lower panels) are similar to those in Figure \ref{fig:mom5} from the SDSS data with the cut (black points)

\end{document}